\documentclass[twocolumn]{aastex62}

\usepackage{txfonts}
\usepackage{graphicx,bm}
\usepackage{color,ulem}
\begin{document}

\shorttitle{The $\nu$ process}
\shortauthors{Sieverding et al.}

\title{The $\nu$-process with Fully Time-dependent Supernova Neutrino
  Emission Spectra} 

\author{A. Sieverding}
\affiliation{School of Physics and Astronomy,
      University of Minnesota, Minneapolis, MN 55455, USA} 
\affiliation{GSI Helmholtzzentrum f{\"u}r
 Schwerionenforschung, Planckstra{\ss}e 1, D-64291 Darmstadt, Germany}
\affiliation{Institut f{\"u}r Kernphysik
 (Theoriezentrum), Technische Universit{\"a}t Darmstadt,
  Schlossgartenstra{\ss}e 2, D-64289 Darmstadt, Germany}
\author{K.~Langanke}
\affiliation{GSI Helmholtzzentrum f{\"u}r
 Schwerionenforschung, Planckstra{\ss}e 1, 64291 Darmstadt, Germany}
\affiliation{Institut f{\"u}r Kernphysik
 (Theoriezentrum), Technische Universit{\"a}t Darmstadt,
  Schlossgartenstra{\ss}e 2, D-64289 Darmstadt, Germany}

\author{G. Mart\'inez-Pinedo}
\affiliation{GSI Helmholtzzentrum f{\"u}r
 Schwerionenforschung, Planckstra{\ss}e 1, 64291 Darmstadt, Germany}
\affiliation{Institut f{\"u}r Kernphysik
 (Theoriezentrum), Technische Universit{\"a}t Darmstadt,
  Schlossgartenstra{\ss}e 2, D-64289 Darmstadt, Germany}

\author{R.~Bollig}
\affiliation{Max-Planck-Institut f\"ur Astrophysik,
  Karl-Schwarzschild-Stra{\ss}e 1, D-85748 Garching, Germany} 

\author{H.-T.~Janka}
\affiliation{Max-Planck-Institut f\"ur Astrophysik,
  Karl-Schwarzschild-Stra{\ss}e 1, D-85748 Garching, Germany} 

\author{A.~Heger}
\affiliation{Monash Centre for Astrophysics, School of Physics and
  Astronomy, Monash University, VIC 3800, Australia} 
\affiliation{Tsung-Dao Lee Institute, Shanghai 200240, China}
\affiliation{The Joint Institute for Nuclear Astrophysics, Michigan
  State University, East Lansing, MI 48824, USA } 

\begin{abstract}
  The neutrino process that occurs in the outer stellar shells
  during a supernova explosion and involves neutrino-nucleus
  reactions produces a range of rare, stable, and radioactive
  isotopes. We improve previous $\nu$-process studies by using, for the
  first time, the time-dependent neutrino emission spectra, as
  predicted from supernova simulations, rather than a simplified
  parametric description modeled after the neutron star cooling
  phase. In particular, our calculations use time-dependent neutrino
  spectra for all neutrino species, consider their deviation from a
  Fermi-Dirac distribution and account for the neutrino emission from
  the neutrino burst and accretion phases. We find that the
  time-dependent treatment of the neutrino emission spectra results in
  higher yields for the selected nuclei produced by the $\nu$~process as
  compared to previous studies and also compared to the approximation of
  assuming constant neutrino energies corresponding to the
  time-averaged mean energy radiated in each species. The effect is
  largest for nuclides produced by charged-current reactions. Our
  results reflect the dynamical competition between neutrino-induced
  reactions and the effect of the shock passage through the star. By
  varying the neutrino burst luminosity and the duration of the
  accretion phase, we study the impact of these early emission phases
  and their uncertainties on the $\nu$-process nucleosynthesis.  We
  find that the deviation of the neutrino spectra from a Fermi-Dirac
  distribution calculated in supernova simulations has a negligible
  effect on the $\nu$-process yields.
\end{abstract}


\section{Introduction}
\label{sec:intro}

Supernova explosions are not only among the brightest observable
events in the universe, they are also the key mechanism to allow the
products of stellar nucleosynthesis to contribute to the chemical
enrichment of the galaxy.  Even though the majority of the nuclei that
constitute the final ejecta of such an explosion are already formed
during the hydrostatic burning phases, the explosion itself leaves an
imprint on the final composition.  Explosive Si, O, and Ne burning are
important for the production of the elements between carbon and iron.
Furthermore, this hot, shock-heated environment allows the production
of light and heavy $p$~nuclei by the
$\gamma$~process~\citep{Woosley.Heger.Weaver:2002,Arnould.Goriely:2003}. 

Supernova nucleosynthesis has been studied in great detail in the last
few decades and current supernova models are very successful in explaining
the solar abundances not only of the elements up to iron, but also of
a range of heavier nuclei produced in the weak $s$~process, the
$\nu p$~process, and the $\gamma$~process
\citep{Froehlich.Martinez-Pinedo.ea:2006,Woosley.Heger:2007,Sukhbold.Ertl.ea:2016,NuGrid,Travaglio.Rauscher.ea:2018,Prantzos.Abia.ea:2018,Wanajo.Mueller.ea:2018,Curtis.Ebinger.ea:2019}. 

\begin{figure*}
 \includegraphics[width=\linewidth]{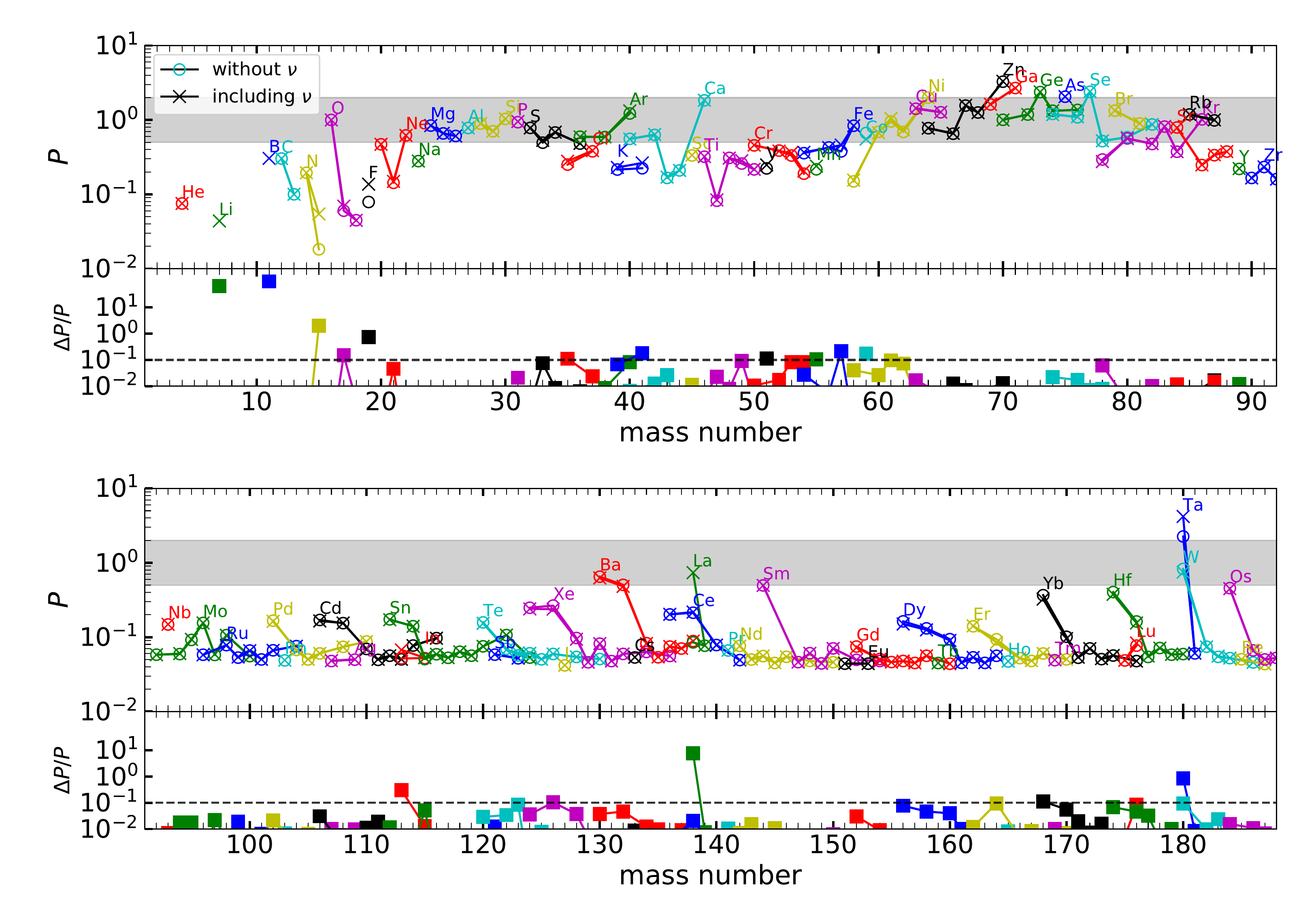}
 \caption{\label{fig:prod_factors_all}The top panels show production factors
	$P$ normalized to $^{16}$O, as defined in Equation
	(\ref{eq:prod_factor}), comparing calculations for a $27\,M_\odot$
	model with and without including neutrino-induced reactions. The gray
	area indicates a factor two around $P=1$. The bottom panels show the
	relative difference between the the calculations with and without
	including neutrinos. The dashed horizontal line indicates a difference
	of $10\,\%$. The isotopes most affected by the $\nu$~process are
	$^7$Li,$^{11}$B,$^{15}$N,$^{19}$F,$^{138}$La and $^{180}$Ta. Other
	isotopes are only affected at the $10\,\%$ level or less. The results
	including the $\nu$~process shown here, follow Approach~1a, which is
	detailed in \S\ref{sec:sim_signal} and Table \ref{tab:approaches}. Note
	that the the production factors for $^7$Li and $^{11}$B without the
	$\nu$~process are less than $10^{-2}$. }
\end{figure*}

In addition to initiating the supernova explosions in the neutrino
driven mechanism, neutrinos also affect the composition of the ejecta
directly.  By now, a range of extensive studies have been performed
that also include the effects of neutrino-induced reactions,
summarized as the $\nu$~process.  Electron neutrino and antineutrino
captures as well as neutral-current spallation reactions involving
all neutrino species, have been shown to be able to contribute to the
production of $^{7}$Li, $^{11}$B, $^{15}$N, $^{19}$F, $^{138}$La, and
$^{180}$Ta~\citep{Woosley.Hartmann.ea:1990,Heger.Kolbe.ea:2005}.
 For illustration,
Figure \ref{fig:prod_factors_all} compares the production factors of stable
isotopes from the supernova yields of the $27\,M_\odot$ stellar model studied
here with and without neutrinos, illustrating the production of the six
$\nu$-process nuclei.  The light elements $^{7}$Li and $^{11}$B do not survive
most stellar processes, and, in order to explain the observed abundances with models
of chemical evolution, contributions from both the $\nu$~process and galactic
cosmic rays are required \citep{Prantzos:2012}.  The production of $^{15}$N in
supernovae is rather small and  recent observations indicate that massive stars
are not the main contributors to the production of $^{19}$F
\citep{Joensson.Ryde.ea:2017}, even though the contribution of supernovae to
those isotopes are relevant in the context of galactic chemical evolution
\citep{Kobayashi.Izutani.ea:2011}. The heavy nuclei $^{138}$La and $^{180}$Ta
both can also be produced by the $\gamma$~process in
Type~Ia supernovae \citep{Travaglio.Rauscher.ea:2018}.  In addition to these
rare, stable isotopes, the $\nu$~process also contributes to the production of
a range of radioactive isotopes, such as $^{22}$Na, $^{26}$Al, $^{36}$Cl
\citep{Sieverding.Martinez.ea:2018}. The production of $^{92}$Nb and $^{98}$Tc
have also been studied in detail by \citet{Cheoun.Ha.ea:2012} and
within
$\nu$-process studies, the effects of MSW flavor oscillations
\citep{Yoshida.Suzuki.ea:2008} and collective oscillations
\citep{Wu.Qian.ea:2015} have also been explored.

Since neutrinos are expected to play a crucial role in successful
supernova explosions, there is great interest in constraints on the
neutrino emission characteristics.  This has been a major motivation
to conceive and improve neutrino detectors that can provide insights
into the supernova mechanism in case an event occurs close enough to
Earth \citep[see][for a review]{Scholberg:2012}.  The $\nu$~process
establishes a direct connection between the production of
individual isotopes and supernova neutrinos. This allows in principle
to constrain the neutrino spectra with nucleosynthesis arguments as
demonstrated by
\citet{Yoshida.Kajino.ea:2005,Yoshida.Kajino.b.ea:2006}, even though the
uncertainties of nuclear and neutrino physics require such arguments
to be taken with caution
\citep{Austin.Heger.Tur:2011,Wu.Qian.ea:2015}.

In addition to the uncertainties mentioned above, such constraints on
the supernova neutrino energies derived from nucleosynthesis arguments
suffer from two conceptual caveats. First of all, they are based on
comparisons to the solar system composition, which is not the result
of a single supernova explosion but involves contributions from a
multitude of events over an incompletely known history of our
galaxy. Therefore, large scale statistical sampling of models is
necessary to draw conclusions.  Another limiting aspect are the strong
simplifications that are made in the modeling of the neutrino
emission. In this paper, we improve this description and quantify the
impact of the latter aspect.  The $\nu$~process has so far mostly been
included in the same parameterization as originally suggested by
\citet{Woosley.Hartmann.ea:1990}, assuming neutrino luminosities that
decrease exponentially with time as
\begin{equation}
\label{eq:param_lum}
 L_\nu(t)=L_0 \,e^{-t/\tau_\nu}
\end{equation}
and Fermi-Dirac spectra with constant average energies
$\langle E_\nu \rangle$ for the different neutrino species.  The
values adopted for the average energies have been revised several
times as supernova simulations have advanced, but the approach has
remained the same \citep{Heger.Kolbe.ea:2005,Yoshida.Suzuki.ea:2008,Banerjee.Qian.ea:2016,Sieverding.Martinez.ea:2018}.

This approach has been justified by the lack of more reliable data
from either observations or from simulations on the details of the
neutrino emission spectra and their time evolution.  The field of
supernova simulations has matured significantly during the last
decades and, in particular, it has arrived at the point that
calculations of different groups with different numerical methods
obtain very similar results if they use the same initial conditions
and assumptions~\citep{OConnor.Bollig.ea:2018}.  This motivates the
present study of the impact of the details of the expected supernova
neutrino signal on the $\nu$~process.

Recent years have brought a deeper understanding of the supernova
mechanism, including, in particular, the role of multidimensional
effects~\citep{Mueller:2015,Janka.Melson.ea:2016,OConnor.Couch:2018,Radice.Abdikamalov.ea:2018,Burrows.Vartanyan.ea:2018}.
While not all aspects of the supernova neutrino emission are
quantitatively agreed upon, the main features have been understood on
a qualitative level that even allows analytic models to achieve
reasonable agreement with numerical
simulations~\citep{Janka:2001,Mueller.ea:2016}.  Three major phases of
neutrino emission from the core of a collapsing star can be
distinguished.

\begin{enumerate}
\item First, there is a very luminous deleptonization outburst of
  electron neutrinos that emerges as the initial bounce shock
  dissociates nuclei into free protons and neutrons and thins out the
  material enough for it to become transparent to the neutrinos
  produced mostly by electron captures on free protons.
\item As the bounce shock stalls, accreting material sheds its
  gravitational binding energy by vigorous emission of all flavors of
  neutrinos produced mostly by thermal processes.  During this phase,
  the diffusive emission from the core is increased by a dynamic
  component that depends on the mass accretion rate and thus on the
  progenitor structure.  The duration of this phase is determined by
  the conditions of shock revival and can be prolonged by
  multidimensional fluid flows.
\item Kelvin-Helmholtz cooling of the nascent neutron star is
  accompanied by strong neutrino emission for about 10~s. Due to the
  long duration, the majority of the neutrinos are emitted during this
  phase, though at lower energies than in the previous phases.
\end{enumerate}

In previous studies of the $\nu$~process
\citep{Woosley.Hartmann.ea:1990,Heger.Kolbe.ea:2005,Yoshida.Suzuki.ea:2008,Sieverding.Martinez.ea:2018}
it is assumed, that only the thermally produced neutrinos from the
neutron star cooling phase are relevant for the nucleosynthesis. In
the cooling phase (Phase~3), all neutrino flavors are produced equally and the
total luminosity is distributed equally among the flavors.  The
$\nu$~process operates more efficiently at stellar radii where
material is not heated to such large temperatures that thermonuclear
reactions modify substantially the composition. Under these
assumptions, the model-dependent time evolution of the neutrino luminosity and energy are of minor importance and the integrated neutrino emission properties such as
the total energy emitted in neutrinos and an estimate for typical neutrino spectra are 
sufficient in order to estimate the nucleosynthesis. The total neutrino energy can, to some
extent, be constrained by the difference of the gravitational binding
energy of the stellar iron core and the final remnant under the
assumption that all the neutrinos are produced during the
proto-neutron star cooling.  This leads to the commonly-used value of
a total energy of $E_{\nu,\text{tot}}=3\times 10^{53}$~erg 
corresponding roughly to the gravitational binding energy of a neutron
star \citep[e.g.][]{Cooperstein:1988}.

\begin{table*}[htb]
  \begin{ruledtabular}
    \caption{Overview of the four different approaches applied in this
      manuscript for the description of the $\nu$-process
      nucleosynthesis. The table lists how the time dependence of the
      average neutrino energies and pinching parameter has been
      treated individually for all neutrino species ($\nu_{e}$,
      $\bar{\nu}_{e}$, $\nu_x$, $\bar{\nu}_x$) based on the data from
      the simulation of \citet{Mirizzi.Tamborra.ea:2016}. All
      approaches consider neutrino fluxes from the beginning of the
      nucleosynthesis calculations but use information from different
      neutrino emission phases. Only Approach~1a considers the full
      time dependence of the neutrino emission. Approaches~2 and~3 use 
      Equation~\ref{eq:param_lum} with constant average neutrino energies.
      Approach 3 neglects
      the burst and accretion phases for the determination of the
      time-averaged energies.  $\alpha = 2.3$ indicates that the
    spectra of Equation~(\ref{eq:alpha_fit}) are used with this constant
    value for $\alpha$ \tablenotemark{a}. \label{tab:approaches}}
    \begin{tabular}{lcccccc}
      & \multicolumn{6}{c}{Neutrino emission phases} \\
      \cline{2-7}
      Approaches & \multicolumn{2}{c}{Neutrino burst} &
      \multicolumn{2}{c}{Accretion phase} &
      \multicolumn{2}{c}{Kelvin-Helmholtz cooling} \\
      \cline{2-3}\cline{4-5}\cline{6-7}
      & $\langle E_\nu \rangle $ & pinching ($\alpha$) & $\langle E_\nu
     \rangle $ & pinching ($\alpha$)  & $\langle E_\nu \rangle $ &
     pinching ($\alpha$) \\ 
      \hline
      1a\tablenotemark{*} & $t$-dependent & $t$-dependent & $t$-dependent & $t$-dependent & $t$-dependent & $t$-dependent \\
      1b & $t$-dependent & $\alpha=2.3$ & $t$-dependent & $\alpha=2.3$ & $t$-dependent & $\alpha=2.3$ \\
       2\tablenotemark{**} & $t$-independent & $\alpha=2.3$ & $t$-independent & $\alpha=2.3$ &  $t$-independent &  $\alpha=2.3$ \\
       3 &  - & - & - & - &  $t$-independent & $\alpha=2.3$  \\
    \end{tabular}
    \tablenotetext{a}{Even if the analytical form of the $\alpha=2.3$
      distribution and of a Fermi-Dirac spectrum are different they
      are indistinguishable in practical numerical applications}
    \tablenotetext{*}{Full spectral information from neutrino
      radiation transport simulations including time dependent
      luminosities, average energies and pinching parameter}
    \tablenotetext{**}{ Assumes the same time independent average
    energy for the three different phases, but all phases are taken into account for the determination of the that energy according to Equation~(\ref{eq:average_def})} 
  \end{ruledtabular}
\end{table*}

In this paper, we test the validity of this approach by performing
$\nu$-process nucleosynthesis studies with $4$ different treatments of
the neutrino emission. In our \textbf{Approaches 1a} and \textbf{1b},
we consider time-dependent neutrino luminosities and average energies
for all neutrino species, taken from a supernova simulation. This
study does not only take the proper time dependence of the neutrino
emission into account, it also considers, for the first time, the
impact of Phases~1 and~2 (burst and accretion, as defined above).
\textbf{Approaches 1a} and \textbf{1b} differ in the treatment of the
neutrino spectra. In \textbf{Approach 1a} we also account for pinched spectra, i.e., 
deviations of the neutrino emission spectra from a Fermi-Dirac
spectrum with zero chemical potential (in the following referred to as
FD spectra). In order to disentangle the effects of pinched neutrino spectra, \textbf{Approach 1b}
assumes FD-like spectra for the emitted neutrinos, as this has been
the case in previous studies of neutrino nucleosynthesis.
\textbf{Approach 1a} represents our full improvement to neutrino
nucleosynthesis, considering the time dependence of the neutrino
emission and including pinched spectra. The calculations performed in \textbf{Approach 1a} are also
confronted with two studies performed in analogy to previous studies,
i.e., with constant average neutrino energies and assuming FD-like
neutrino spectra.  These two approaches differ in
the way the time average of the neutrino energies is determined.
\textbf{Approach 2} is consistent with the first calculation; i.e.,
the average is performed over the complete neutrino emission spectrum,
including Phases~1 and~2. Thus, it allows to study the sensitivity of the nucleosynthesis results to the time evolution of the neutrino emission.
\textbf{Approach 3} more closely follows
previous studies, by considering in the average only neutrinos emitted
during the cooling phase. It neglects the neutrino emission Phases 1 and 2 and thus allows to pin down their effect on the nucleosynthesis.

In Approaches~2 and~3, total energies
$E_{\nu,\text{tot}}$ are obtained from the simulation for $\nu_{e}$ and
$\bar{\nu}_{e}$ and the heavy flavors $\nu_x$ and $\bar{\nu}_x$
separately, but they include, on the one hand, the whole duration of the emission (Approach~2) and, on
the other hand, only the Kelvin-Helmholtz cooling phase (Approach~3).
The parameterization of the neutrino luminosities is then adjusted to the total energies,
and the average energies of the various neutrino species are
obtained as
$\langle E_\nu \rangle =E_{\nu,\text{tot}}/N_{\nu,\text{tot}} $, such that
the integrated number of emitted neutrinos
$N_{\nu,\text{tot}} $ equals the included phases of the supernova simulation.
With this definition of time-averaged energies the
total number of neutrinos and the energy are the same in Approaches~1
and~2. Table \ref{tab:approaches} gives an overview of the four approaches used in this paper.

More details about the different approaches and of our calculations are given in \S\ref{sec:sim_signal} together with a discussion of the simulation data we use.  The impact on the
nucleosynthesis is presented in \S\ref{sec:impact} with further
details on the role of pinched neutrino spectra in \S\ref{sec:alphas}.
Finally, the sensitivity to variations of the accretion time and the
neutrino burst luminosity are explored in \S\ref{sec:sensitivity}.

\section{Neutrino signal from supernova simulations}
\label{sec:sim_signal}

In our $\nu$-process calculations, we combine a piston-driven explosion
model as in \citet{Woosley.Weaver:1995}, tuned to produce an explosion
energy of $1.2 \times 10^{51}$~erg, with neutrino
emission data from a numerical simulation (discussed in more detail
below) and a large nuclear network.  The neutrino emission is
sensitive to the detailed dynamics of the high-density matter at the
stellar core. As the $\nu$~process occurs at radii larger than
$1000$~km, the neutrino emission processes and the propagation of the
shock through the mantle can be treated as decoupled processes.

For the $\nu$~process we need to take into account neutrino-induced charged-
and neutral-current reactions that affect the nuclear
composition~\citep{Kolbe.Langanke.ea:2003}.  In the charged-current
reactions, the neutrino (antineutrino) transforms a neutron (proton)
in the nucleus into a proton (neutron).  At the supernova neutrino
energies involved, this process can only occur for electron neutrinos
and antineutrinos. It might, however, be accompanied by the emission
of a single or multiple light particles (proton, neutron, $\alpha$
particle).  Neutral-current reactions can be initiated by all neutrino
flavors. For the $\nu$~process we are interested in neutrino-induced
spallation reactions in which the neutrino excites the nucleus above
(single or multiple) particle emission thresholds so that the excited
nuclear level can decay by particle emission, changing the matter
composition.  The relevant neutrino-induced partial reaction cross
sections are described in a two-step process, following
\citet{Kolbe.Langanke.ea:1992,Langanke.Vogel.Kolbe:1996,Balasi.Langanke.ea:2015,Huther:2014}. In
the first step, the neutrino-induced nuclear excitation function is
calculated. This step is usually performed within the framework of the
Random Phase Approximation (RPA), but for selected nuclei these
calculations are based on large-scale shell model calculations or on
experimental data on Gamow-Teller strength
functions~\citep[see][]{Sieverding.Martinez.ea:2018}.  The second
step, i.e., the decay of the excited nuclear levels, is described
within the statistical model, including also multi-particle emission
channels~\citep{Huther:2014,Sieverding.Martinez.ea:2018}.

The neutrino-induced partial differential cross sections are
incorporated into our nuclear reaction network, which evolves the
abundances, $Y_i(t)$, of 1988 nuclear species connected by the
thermonuclear reaction rates from the JINA REACLIB
database~\citep{Cyburt.Amthor.ea:2010} and $\beta$ decays from the
NUBASE compilation of experimentally determined
values~\citep{Audi2017} where available and otherwise from the
theoretical predictions by~\citet{Moeller.Pfeiffer.ea:2003}. In the equations of the
nuclear reaction network, the
neutrino-nucleus reactions enter as additional terms in the form
\begin{equation}
  \label{eq:fold} 
  \frac{\partial Y_i(t)}{\partial t} = \langle \sigma_\nu \rangle_{i,j}
  (t)\, \phi_\nu (t)\, Y_j(t),
\end{equation}
with
\begin{equation}
  \label{eq:integ_csect}
  \langle \sigma_\nu \rangle_{i,j} (t) = \int\limits_0^\infty
  \sigma_{i,j}(E_\nu) n_\nu(E_\nu,t)\; dE_\nu,
\end{equation}
where
$\phi_\nu (t)=L_\nu(t)/[4\pi\, r^2(t)\,\langle E_\nu \rangle(t)]$ is
the neutrino number flux at radius $r$ and $\sigma_{i,j}(E_\nu)$ is the
energy-dependent neutrino-nucleus cross section. The indices $j,i$
refer to the struck nucleus and the larger nuclear fragment in the
final channel, respectively, and $n_\nu(E_\nu,t)$ is the normalized
neutrino spectrum at time $t$. Appropriate equations are included for
the light particle produced (proton, neutron, $\alpha$-particle) by
the neutrino-induced reactions and for the struck nucleus where its
abundance, $Y_j$, is decreased by the neutrino-nucleus reaction.

In our studies we use the neutrino-nucleus reaction cross sections of
\citet{Sieverding.Martinez.ea:2018}, except for the reactions on
$^4$He and $^{12}$C, which are taken from shell model calculations by \citet{Yoshida.Suzuki.ea:2008}.  The neutrino-induced
reaction rates $\langle \sigma_\nu \rangle_{i,j}$ are then calculated
with the appropriate neutrino emission data, considering
time-dependent (Approaches~1a and 1b) or constant (Approaches 2 and 3)
average neutrino energies.

\begin{figure}[htb]
   \centering
   \includegraphics[width=\linewidth]{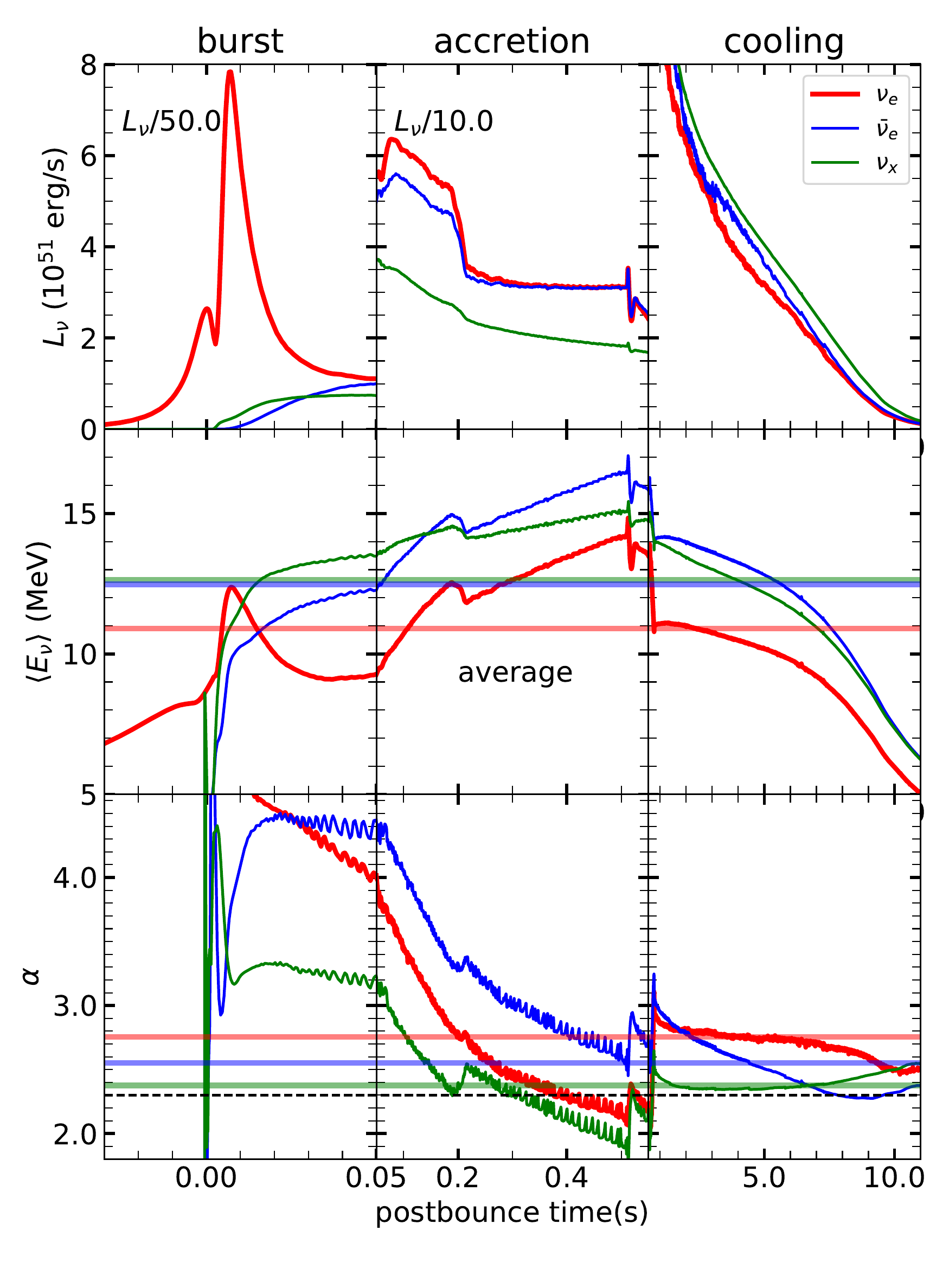}
   \caption{Time evolution of neutrino luminosities and energies from
     a one-dimensional, artificially triggered supernova simulation
     \citep{Mirizzi.Tamborra.ea:2016}. The calculation uses the SFHo
     equation of state \citep{SFH}.  The top panels show the
     luminosities and the middle panels the average neutrino energies.
     The horizontal lines in the middle panel indicate the
     representative energies as defined by
     Equation~(\ref{eq:average_def}). During the accretion period, the
     actual neutrino energies are substantially higher than the
     representative values. The bottom panel shows the value of
     $\alpha$ as defined in Equation~(\ref{eq:alpha_fit}) that describes the
     pinching of the neutrino spectra. \label{fig:sim_signal_SFHo}}
\end{figure}

Following~\citet{Keil.Raffelt.ea:2003} and
\citet{Tamborra.Mueller.ea:2012}, the instantaneous normalized neutrino
spectra are represented by a quasi-thermal distribution:

\begin{equation}
  \label{eq:alpha_fit}
    n_\nu (E) \equiv \left(\frac{\alpha+1}{\langle
      E\rangle}\right)^{\alpha+1}\frac{E^\alpha}{\Gamma(\alpha+1)}\exp\left(-\frac{(\alpha+1)
      E}{\langle E \rangle}\right),
\end{equation}
with the gamma function $\Gamma$.
The parameter $\alpha$ can be obtained from the second moment of the neutrino
spectra:  
\begin{equation}
 \frac{\langle E_\nu^2 \rangle }{\langle E_\nu \rangle
   ^2}=\frac{\alpha+2}{\alpha+1}. 
\end{equation}
A value of $\alpha=2.0$ corresponds to the limit of a
Maxwell-Boltzmann distribution, whereas $\alpha=2.3$ very closely
matches a Fermi-Dirac spectrum with zero chemical potential,

\begin{equation}
  \label{eq:fdspectra}
  n_{\text{FD}}(E) = \frac{2}{3\zeta(3) T_\nu^3} \frac{E^2}{\exp(E/T_\nu)+1},
\end{equation}
with the neutrino temperature $T_\nu$ related to the average
neutrino energy $T_\nu = 180\zeta(3) \langle E_\nu
\rangle/(7\pi^4)$, where $\zeta$ is the Riemann zeta-function. Values $\alpha>2.3$ account for increasingly
pinched spectra, whereas $\alpha<2.3$ corresponds to an
``antipinching'' in which high-energy neutrinos are more likely to be
encountered than in the equilibrium Fermi-Dirac distribution. 
 It has been found
already by early calculations
\citep{Janka.Hillebrandt:1989,Giovanoni.Ellison.ea:1989,Myra.Burrows.ea:1990}
that the neutrino spectra emitted from a supernova explosion tend to be 
pinched and also modern calculations
\citep{Keil.Raffelt.ea:2003,Tamborra.Mueller.ea:2012,Mirizzi.Tamborra.ea:2016} show the same trend (see
Figure~\ref{fig:sim_signal_SFHo}).
With relevance to neutrino nucleosynthesis we note that, in general,
pinched spectra result in reduced folded neutrino-nucleus cross
sections, $\langle \sigma_\nu \rangle$,  compared to those obtained with FD spectra, as the number of
high-energy neutrinos is reduced, while on the other hand anti-pinched
spectra yield larger cross sections. Neutrino-nucleus cross sections
for different forms of neutrino spectra are given
in~\citep{Kolbe.Langanke.ea:1992,Langanke.Kolbe:2001}.

We use neutrino luminosities and spectra from a one-dimensional,
artificially exploded supernova
simulation~\citep{Mirizzi.Tamborra.ea:2016}. It included a detailed
treatment of neutrino transport with a two-moment scheme and a
variable Eddington factor closure derived from a model-Boltzmann
equation~\citep{Rampp.Janka:2002}, which constitutes an efficient
numerical method to solve the integro-differential transport problem
and gives results in good agreement with a direct solution of the
Boltzmann equations~\citep{Liebendoerfer.Rampp.ea:2005}.

This model has also been used by~\citet{Bartl.Bollig.ea:2016} to study
the effects of an improved treatment of nucleon-nucleon bremsstrahlung
on the proto-neutron star cooling phase.  We remark that, at this point,
we have to resort to artificially triggered one-dimensional supernova
explosions for our $\nu$-process studies because self-consistent 3D
models are computationally very expensive and are therefore usually
not run long enough to cover the whole proto-neutron star~(PNS)
cooling phase which is important for the $\nu$~process.  The
one-dimensional simulation for a 27~M$_\odot$ progenitor used here includes
effects of PNS convection treated with a mixing-length description.
It is suitable for our purposes because it exhibits all the relevant
features of the neutrino signal that are qualitatively also found in
self-consistent multidimensional
models~\citep{Mirizzi.Tamborra.ea:2016}, and we consistently
combine it with the 27~M$_\odot$ progenitor model
from~\citet{Woosley.Heger:2007}, from which we take the pre-supernova
abundances and structure. 

Figure~\ref{fig:sim_signal_SFHo} shows the neutrino luminosities and
spectral properties as obtained in the simulation
of~\citet{Mirizzi.Tamborra.ea:2016}.  In particular, the figure
distinguishes the three different emission phases as defined above.
The two upper panels show the luminosities and the average neutrino
energies, while the bottom panel shows the $\alpha$ parameter that
characterizes the deviation from a FD spectrum.  We note that the
electron neutrino spectra are strongly pinched during the burst
phase. This is related to the fact that the electron neutrinos with
higher energies are affected by charged-current reactions even after
$\nu_{e}$-neutrinos with average energies are decoupled from
matter~\citep{Giovanoni.Ellison.ea:1989}. During the accretion phase
there is a relatively strong variation in the neutrino spectral forms,
including short periods of anti-pinched spectra for $\nu_{e}$ and
$\nu_x$. During the cooling phase, the $\nu_x$ spectrum closely resembles
a FD~spectrum, while $\nu_{e}$ and $\bar{\nu}_{e}$ spectra are
slightly pinched.  $\bar{\nu}_x$ behave very similar to $\nu_x$ as shown by \citet{Mirizzi.Tamborra.ea:2016} and are omitted in Figure \ref{fig:sim_signal_SFHo}.

Our approach improves previous $\nu$-process studies in four important
aspects: (\emph{i}) It considers the electron neutrino burst (Phase~1)
which occurs during the first $10\,\mathrm{ms}$ after bounce and is associated
with the shock breakout from the neutrino trapping
regime~\citep[e.g.][]{Janka.Langanke:2007}. The burst alone contains
$5\,\%$ of the total energy that is emitted in electron neutrinos. The
burst phase was not considered in previous studies. (\emph{ii}) Our
study includes the prolonged phase of accretion (Phase~2) as material
falls through the stalled shock. In
\citet{Mirizzi.Tamborra.ea:2016}, the explosion has been artificially
initiated at 0.5~s after the bounce, effectively ending the accretion
phase, since in a one-dimensional model matter cannot pass around the
expanding material. We note that the duration of the accretion phase
is somewhat uncertain and that the longer this period lasts, the more
effectively neutrinos can contribute to the nucleosynthesis. We study
the impact of the length of the accretion phase on the neutrino
nucleosynthesis in \S\ref{sec:sensitivity}. The accretion phase has
not been included in previous studies. (\emph{iii}) Rather than using
the parametric form of Equation~(\ref{eq:param_lum}), we consider the time-dependence of the
luminosities for the individual neutrino species as given by
\citet{Mirizzi.Tamborra.ea:2016}. We note, however, that the
luminosities during the cooling phase, which starts around
$500\,\mathrm{ms}$ after bounce, are relatively well described by an
exponential decline $L_{\nu}\propto e^{-t/\tau}$ with $\tau=3\,\mathrm{s}$, as
adopted in previous studies (and in our approaches 2 and
3). (\emph{iv}) 
Since the supernova simulation treats the neutrino transport with multiple energy groups,
it does not only contain
information about neutrino fluxes and the average energies but it
also provides information about the spectral shape, expressed in terms of $\langle E^2 \rangle$, that is related to $\alpha$ by Equation~(\ref{eq:alpha_fit})

\begin{table}[htb]
  \begin{ruledtabular}
    \caption{Cross sections for the reactions for which the spectral shape was taken into account\label{tab:alpha_reactions}}
    \begin{tabular}{lccc}
    Reaction & \multicolumn{3}{c}{Cross section in
               $10^{-42}\,\mathrm{cm}^2$} \\
      \cline{2-4}
             &  $\alpha=2.3$  & $\alpha=2.5$ & $\alpha=2.8$ \\ \hline
   $^{4}$He$(\nu,\nu'p)^{3}$He       &  $  1.00\times 10^{-2} $ & $  8.31\times 10^{-3}  $ & $  6.38\times 10^{-3} $ \\ 
   $^{4}$He$(\nu,\nu'n)^{3}$He       &  $  9.26\times 10^{-3} $ & $  7.68\times 10^{-3}  $ & $  5.88\times 10^{-3} $ \\
   $^{12}$C$(\nu,\nu'p)^{11}$B       &  $  3.44\times 10^{-2} $ & $  2.94\times 10^{-2}  $ & $  2.35\times 10^{-2} $ \\
   $^{12}$C$(\nu,\nu'n)^{11}$C       &  $  9.86\times 10^{-3} $ & $  8.30\times 10^{-3}  $ & $  6.51\times 10^{-3} $ \\
   $^{16}$O$(\nu,\nu'p)^{15}$N       &  $  6.01\times 10^{-2} $ & $  5.14\times 10^{-2}  $ & $  4.14\times 10^{-2} $ \\
   $^{16}$O$(\nu,\nu'n)^{15}$O       &  $  1.51\times 10^{-2} $ & $  1.28\times 10^{-2}  $ & $  1.01\times 10^{-2} $ \\
   $^{16}$O$(\nu,\nu' \alpha p)^{11}$B&  $  1.22\times 10^{-5} $ & $  9.21\times 10^{-6}  $ & $  6.18\times 10^{-6} $ \\        
   $^{16}$O$(\nu,\nu' \alpha n)^{11}$C&  $  2.73\times 10^{-5} $ & $  2.08\times 10^{-5}  $ & $  1.42\times 10^{-5} $ \\        
   $^{20}$Ne$(\nu,\nu'p)^{19}$F      &  $  3.15\times 10^{-3} $ & $  2.66\times 10^{-3}  $ & $  2.11\times 10^{-3} $ \\
   $^{20}$Ne$(\nu,\nu'n)^{19}$Ne     &  $  8.29\times 10^{-2} $ & $  7.22\times 10^{-2}  $ & $  5.94\times 10^{-2} $ \\
   $^{12}$C$(\nu_{e},e^{-}p)^{11}$C      &  $  8.58\times 10^{-2} $ & $  7.12\times 10^{-2}  $ & $  5.47\times 10^{-2} $ \\
   $^{12}$C$(\bar{\nu}_{e},e^{+}n)^{11}$B&  $  3.20\times 10^{-4} $ & $  2.59\times 10^{-4}  $ & $  1.91\times 10^{-4} $ \\
   $^{138}$Ba$(\nu_{e},e^{-} )^{138}$La   &  $  6.36\times 10^{1} $ & $  6.20\times 10^{1}  $ & $  5.98\times 10^{1} $ \\
   $^{138}$Ba$(\nu_{e},e^{-}n)^{137}$La  &  $  6.22\times 10^{1} $ & $  5.82\times 10^{1}  $ & $  5.30\times 10^{1} $ \\
   $^{180}$Hf$(\nu_{e},e^{-} )^{180}$Ta   &  $  1.16\times 10^{2} $ & $  1.13\times 10^{2}  $ & $  1.10\times 10^{2} $ \\
   $^{180}$Hf$(\nu_{e},e^{-}n)^{179}$Ta  &  $  7.61\times 10^{1} $ & $  7.22\times 10^{1}  $ & $  6.71\times 10^{1} $ \\
 \end{tabular}
 \tablecomments{Cross sections have been folded with spectra as in
   Equation~(\ref{eq:alpha_fit})  with different values of $\alpha$  and
   $\langle E_\nu \rangle=12.6$~MeV.}  
  \end{ruledtabular}
\end{table}

Including the effects of the spectral shape requires to fold the cross
section with the spectrum according to Equation~(\ref{eq:integ_csect})
for each time step of the calculation. This is computationally
challenging. Therefore, we include the full spectral shape only for a
limited set of neutrino-nucleus reactions that are the most important.
The reactions are listed in Table~\ref{tab:alpha_reactions} and
include the main reactions for the production of $^{7}$Li, $^{11}$B,
$^{15}$N, $^{19}$F, $^{138}$La and $^{180}$Ta. Further details
on those cross-sections can be
found in the Appendix. The production of
$^{7}$Li and $^{11}$B in the He~shell proceeds via the reactions
$^{3}$He$(\alpha,\gamma)^{7}$Be$(\alpha,\gamma)^{11}$C and
$^{3}$H$(\alpha,\gamma)^{7}$Li$(\alpha,\gamma)^{11}$B, initiated by
neutrino-induced reactions on $^4$He to provide $^{3}$He and $^{3}$H,
which are not produced significantly by thermal photons because the
matter temperature is too low.  For all other reactions we assume a
Fermi-Dirac spectrum $n_{\text{FD}}(E_\nu)$ and interpolate the cross
sections from a set of values calculated as a function of the spectral
average energy $\langle E_\nu \rangle (t)$. In this way we account for
the evolution of the neutrino spectra with time as predicted by the
simulation.

In Table~\ref{tab:alpha_reactions} we compare the folded cross sections, $\langle \sigma_\nu \rangle$, for a FD-like spectrum
($\alpha=2.3$) with those obtained with pinched spectra ($\alpha=2.5$
and $2.8$) as they occur for electron neutrinos and electron
antineutrinos in the cooling phase and for $\nu_x$ neutrinos in the
accretion phase, assuming a typical average energy $\langle E_\nu \rangle=12.6$~MeV. The
cross sections decrease with increasing value of $\alpha$,
i.e.\ pinching, but the effect of the pinched spectra on the cross
sections depends also quite sensitively on the reaction thresholds. As
neutral-current induced spallation reactions have in general large
particle emission thresholds, the effect of pinching is sizable here,
as can be seen for the neutral-current reactions on $^4$He, $^{12}$C
and on $^{20}$Ne. However, neutral-current reactions are mainly
induced by $\nu_x$ neutrinos during the cooling phase and their
spectra in this phase resemble FD spectra relatively closely. Hence,
the consideration of pinched spectra, as performed here for the first
time, should not have too strong an impact on the neutrino
nucleosynthesis yields of nuclides which are produced by
neutral-current reactions. On the other hand, charged-current
reactions have smaller threshold energies, which reduces the impact of
the pinched spectra. The results obtained for the $(\nu_{e},e^{-})$
reactions on $^{138}$Ba and $^{180}$Hf are examples. In both cases the
cross section is reduced by a few percent, comparing the cross
sections for the FD-like~spectrum ($\alpha=2.3$) to the values obtained with the pinched
spectrum with $\alpha=2.8$. The effect is somewhat larger, if one compares
charged-current reactions with emission of particles (neutrons for the
cases of $^{138}$Ba and $^{180}$Hf), because an additional threshold has to
be overcome. The effect is more significant for the
$(\nu_{e}, e^{-}p)$ and $(\bar{\nu}_{e},e^{+}n)$ reaction
cross sections on $^{12}$C due to the relatively large threshold
energies involved.  The effects on the nucleosynthesis yields are
explored in more detail in \S\ref{sec:alphas}.

We adjust the time of the neutrino data such that the peak in the
electron neutrino luminosity coincides with the time the piston is
launched.  We assume that the neutrinos travel at the speed of light
and therefore the arrival of the neutrino signal is slightly delayed
for mass shells at larger radii. 

The result of our Approach~1a, which uses the fully time-dependent neutrino data from the simulation, is compared to Approaches~2 and~3 that use
approximations similar to previous neutrino nucleosynthesis
calculations and also assume a constant value of $\alpha=2.3$ for the
reactions listed in table~\ref{tab:alpha_reactions}. In these models average energies
are constant and the neutrino luminosity is described by Equation~(\ref{eq:param_lum}).
$L_0$ is fixed by requiring that the time integrated luminosity gives the same value of the total energy emitted as neutrinos, 
\begin{equation}
\label{eq:tot_energy}
 E_{\nu,\text{tot}}=\int\limits_{t_0}^{t_\mathrm{end}} L_\nu(t)\,\mathrm{d}t\;,
\end{equation}
as obtained with the numerical values of $L_\nu(t)$ provided by the
simulation, which covers the time from $t_0=-0.3$~s (before
bounce) until $t_{\text{end}}=11.2$~s after bounce. 

Based on the simulation, the total energy emitted as neutrinos, $E_{\nu,\text{tot}}$,  is $3.49\times 10^{53}$~erg.
The distribution among the neutrino flavors is almost equal. We find
that $\nu_{e}$ contribute $0.62\times 10^{53}$~erg, $\bar{\nu}_{e}$ amount
to $0.59\times 10^{53}$~erg and the heavy flavors $\mu$ and $\tau$
give $2.27\times 10^{53}$~erg for neutrinos and antineutrinos
together.
In our Approach~2,
the entire neutrino signal is considered to calculate
$E_{\nu,\text{tot}}$. 
Whereas Approach~3 is similar to previous studies in that it takes
into account only the PNS cooling phase starting at $t_0=500$~ms after
bounce, when the explosion was triggered.  This results in a total
energy emitted as neutrinos of $2.53\times 10^{53}$~erg, of which
$0.37\times 10^{53}$~erg are from $\nu_{e}$, $0.40\times 10^{53}$~erg
are from $\bar{\nu}_{e}$ and the remainder from the heavy flavor
neutrinos.

A second integral
quantity of the neutrino signal is the total number of emitted
neutrinos, that we calculate from the simulation data as
\begin{equation}
\label{eq:tot_number}
 N_{\nu,\text{tot}}=\int\limits_{t_0}^{t_\mathrm{end}}
 \frac{L_\nu(t)}{\langle E_\nu \rangle (t)} dt. 
\end{equation}
Again, in our Approach~2,
the entire neutrino signal is considered to calculate  $N_{\nu,\text{tot}}$, while Approach~3 limits the integration to the cooling phase, i.e., $t_0=0.5\,\mathrm{s}$.

With different values of $E_{\nu,\text{tot}}$ and $N_{\nu,\text{tot}}$ 
Approaches~2 and~3 also use different values of the constant, time-averaged neutrino energies calculated as 
\begin{equation}
    \langle E_\nu \rangle= \frac{ E_{\nu,\text{tot}}}{  N_{\nu,\text{tot}} }
    \label{eq:average_def}
\end{equation}

For Approach 2 we obtain 
$\langle E_{\nu_{e}} \rangle =10.9$~MeV, 
$\langle E_{\bar{\nu}_{e}} \rangle =12.6$~MeV, 
$\langle E_{\nu_x} \rangle =11.7$~MeV and 
$\langle E_{\bar{\nu}_x} \rangle=12.6$~MeV.
These are effectively the same average values as the set of ``low''
neutrino energies discussed and adopted
in~\citet{Sieverding.Martinez.ea:2018} except for the electron
neutrinos. 

For Approach~3 we find
$\langle E_{\nu_{e}} \rangle=10.6$~MeV, 
$\langle E_{\bar{\nu}_{e}} \rangle =13.3$~MeV, 
$\langle E_{\nu_x} \rangle=12.5$~MeV and 
$\langle E_{\bar{\nu}_x} \rangle=13.3$~MeV, relatively 
close to the values used by~\citet{Sieverding.Martinez.ea:2018}, i.e.,
$\langle E_{\nu_{e}} \rangle=8.8$~MeV, 
$\langle E_{\bar{\nu}_{e}} \rangle =\langle E_{\nu_x} \rangle=\langle E_{\bar{\nu}_x} \rangle =12.6$~MeV.

In Approaches~2 and~3 we use FD-like neutrino spectra with a constant (time-independent)
average energy, which still differs for $\nu_{e}$, $\bar{\nu}_{e}$,$\nu_x$
and $\bar{\nu}_x$ neutrinos. This ansatz is equivalent to the
assumption of constant time-independent average energies for the
different neutrino species.

Nucleosynthesis studies published prior to 2018 used neutrino emission
spectra with noticeably higher energies, as they were appropriate at
the time they were performed, e.g.,
$\langle E_{\nu_{e}}\rangle=12.6$~MeV,
$\langle E_{\bar{\nu}_{e}} \rangle = 15.8$~MeV,
$\langle E_{\nu_x} \rangle= \langle E_{\bar{\nu}_x} \rangle =18.9$~MeV
in~\citet{Heger.Kolbe.ea:2005}.

\section{Impact of the improved description of the neutrino emission on the $\nu$~process}
\label{sec:impact}

In this section, we report on the results which we obtain in our
nucleosynthesis studies for a 27~M$_\odot$ progenitor star using the
improved neutrino emission description based on the supernova
simulation of \citet{Mirizzi.Tamborra.ea:2016} and defined in the
previous section (Approaches~1a and~1b).  Note that Approach~1a
includes also effects of the pinching of neutrino spectra which are
discussed in detail in \S\ref{sec:alphas}. 
The $27\,M_\odot$
progenitor star used here does not reflect the full picture of neutrino
nucleosynthesis. It has some peculiarities which are, for example, not
found in lower-mass progenitors~\citep{Sieverding.Martinez.ea:2018}
and which we will address below. Our goal here is to explore the
impact of the various improvements which we consider in the
description of the neutrino emission signal. To this end, the results
of our Approach~1b are compared to calculations in which these
improvements were either treated approximately (time-independent
average neutrino energies, Approach~2) or partly ignored (no
consideration of the burst and accretion phases and time-independent
average neutrino energies, Approach~3). As stated above, Approach~3
reflects the spirit of previous studies of neutrino nucleosynthesis.

\begin{table}[htb]
  \begin{ruledtabular}
    \caption{Production factors \label{tab:prodfac_full_signal}}
    \begin{tabular}{l|ccccc}

             &     Appr. 1a & Appr. 1b & Appr. 2 & Appr. 3 &  Literature\\
     Nucleus           &  $\alpha=\alpha(t)$& $\alpha=2.3$&  $\alpha=2.3$ & $\alpha=2.3$ & FD                                                                           \\ \hline
     $^7$Li     & 0.04&  0.04 &  0.03& 0.02  & 0.02 \\
     $^{11}$B   & 0.30& 0.31   & 0.28 & 0.18 & 0.13 \\
     $^{15}$N   & 0.06& 0.05   & 0.05 & 0.04 & 0.04 \\
     $^{19}$F   & 0.12& 0.12   & 0.11 & 0.10 & 0.10 \\
     $^{138}$La & 0.69& 0.74   & 0.66 & 0.41 & 0.44 \\
     $^{180}$Ta$^{\rm m}$ & 1.32 &  1.33  & 1.27 & 1.09  & 1.11  \\
 \end{tabular} 
 \tablecomments{Production factors are normalized to $^{16}$O, see
   Equation~(\ref{eq:prod_factor}), comparing the different approaches
   for the description of the neutrino irradiation discussed in the
   text for the 27~M$_\odot$ model. Only Approach~1a takes the
   spectral shape described by $\alpha(t)$ into account and is
   discussed in \S\ref{sec:alphas}. The column labeled ``Literature''
   gives the results of \citet{Sieverding.Martinez.ea:2018} who assumed
   Fermi-Dirac spectra with vanishing chemical potential for the
   neutrinos.
The results given for  $^{180}$Ta$^{\text{m}}$ assume that 35\% of the
produced $^{180}$Ta at $200\,\mathrm{s}$ after bounce survives in the isomeric state $^{180}$Ta$^{\text{m}}$.} 
\end{ruledtabular}
\end{table}

The results of our nucleosynthesis studies are summarized in
Table~\ref{tab:prodfac_full_signal}.  We list production factors, $P$,
normalized to $^{16}$O, i.e.,

\begin{equation}\label{eq:prod_factor}
  P=[X_*(A,Z)/X_\odot(A,Z)]/[X_*({}^{16}\text{O})/X_\odot({}^{16}\text{O})],
\end{equation}
with solar mass fractions $X_\odot$ from \citet{Lodders:2003} and the
stellar mass fractions $X_*$ obtained from our calculations.  More
recent evaluations of the solar system abundances are
available~\citep{Lodders:2009,Asplund.Grevesse.ea:2009}, but for
consistency within this paper we use the same values that were assumed
for the initial composition of the progenitor model. The ground state
of $^{180}$Ta has too short a $\beta$-decay half-life (8.2~hr) to
contribute to the abundance in the solar system. Only the long-lived
isomeric state $^{180}$Ta$^{\text{m}}$ with a halflife of
$\sim 10^{15}$~yr is still present.  In our calculations, we do not
follow the ground and isomeric states separately, but rather determine
the population of the isomeric state in our reaction network by
adopting the estimate of \citet{Mohr.Kaeppeler.Gallino:2007} and
\citet{Hayakawa.Mohr.ea:2010} that 35\%--39\% of the total $^{180}$Ta
abundance survives in the isomeric state. The yields and production
factors for $^{180}$Ta$^\mathrm{m}$ shown in
Table~\ref{tab:prodfac_full_signal} are 35\% of the calculated
$^{180}$Ta yields.

Before entering the detailed discussion of the effect of our improved
treatment of the neutrino emission, 
we note that, qualitatively, our calculations using the
full neutrino emission signal (Approaches~1a and~1b) confirm the general conclusions drawn in
\citet{Sieverding.Martinez.ea:2018} for the neutrino-induced production of the nuclides  $^7$Li,
$^{11}$B, $^{15}$N, $^{19}$F, $^{138}$La, and $^{180}$Ta. In agreement with previous studies
\citep{Woosley.Hartmann.ea:1990,Heger.Kolbe.ea:2005,Sieverding.Martinez.ea:2018}, using an extensive
nuclear network,
we do not find evidence for other nuclides being significantly produced by the $\nu$~process.
Comparing our results with Approaches~1a and~1b to the more approximate 
Approaches~2 and~3, however, we find some
significant differences that demonstrate the importance
of the improved treatment of the neutrino signal.
Table~\ref{tab:prodfac_full_signal} shows that differences are largest
between Approach~1b and Approach~3.  For Approach~3, which is in the spirit of
the previous studies, the yields are noticeably less than in our
improved study with differences ranging between 16\% for $^{19}$F
to 50\% for $^7$Li.  This shows that the neutrino emission from the
burst and accretion phases needs to be included.  With Approach~2,
considering all three neutrino emission phases, the reduction is
noticeably smaller, but in Approach~2 we have ignored the time dependence of the average
neutrino energies by using constant average energies for the
individual neutrino species. With this approach we still find smaller
nucleosynthesis yields for the $\nu$-process nuclei compared to
Approach~1a but the reduction is on the level of a few percent for all
species. This difference is due to the energy dependence that enters
the neutrino-nucleus cross section via the phase space factor. This
additional energy factor favors the contribution arising from neutrino
energies higher than average relative to those lower than average and is explained 
in more detail below.  We
note that the yields calculated in our Approach~3 agree quite well
with those presented by \citet{Sieverding.Martinez.ea:2018} using the
same progenitor and explosion model. As stated above, Approach~3 is
performed using the same general assumptions in the description of the
neutrino emission signal and the remaining differences between the two calculations
can be traced back to two compensating effects: The time-integrated
neutrino luminosity assumed by \citet{Sieverding.Martinez.ea:2018}, is
slightly larger
than that calculated in Approach~3 ($3\,\times10^{53}\,\mathrm{erg}$
compared to $2.53\,\times 10^{53}\,\mathrm{erg}$, respectively). This
reduces the yields with our Approach~3.  On the other hand, the average energies of the
various neutrino species are slightly smaller in
\citet{Sieverding.Martinez.ea:2018} than in our Approach~3,
increasing our yields. Our Approach~3 results in significantly lower
yields for the $\nu$-process isotopes than obtained in earlier studies~\citep[e.g.,][]{Woosley.Hartmann.ea:1990,Heger.Kolbe.ea:2005,Yoshida.Kajino.ea:2005,Yoshida.Suzuki.ea:2008},
who assumed neutrino spectra with noticeably larger average energies.

\begin{figure}[htb]
   \centering
   \includegraphics[width=\linewidth]{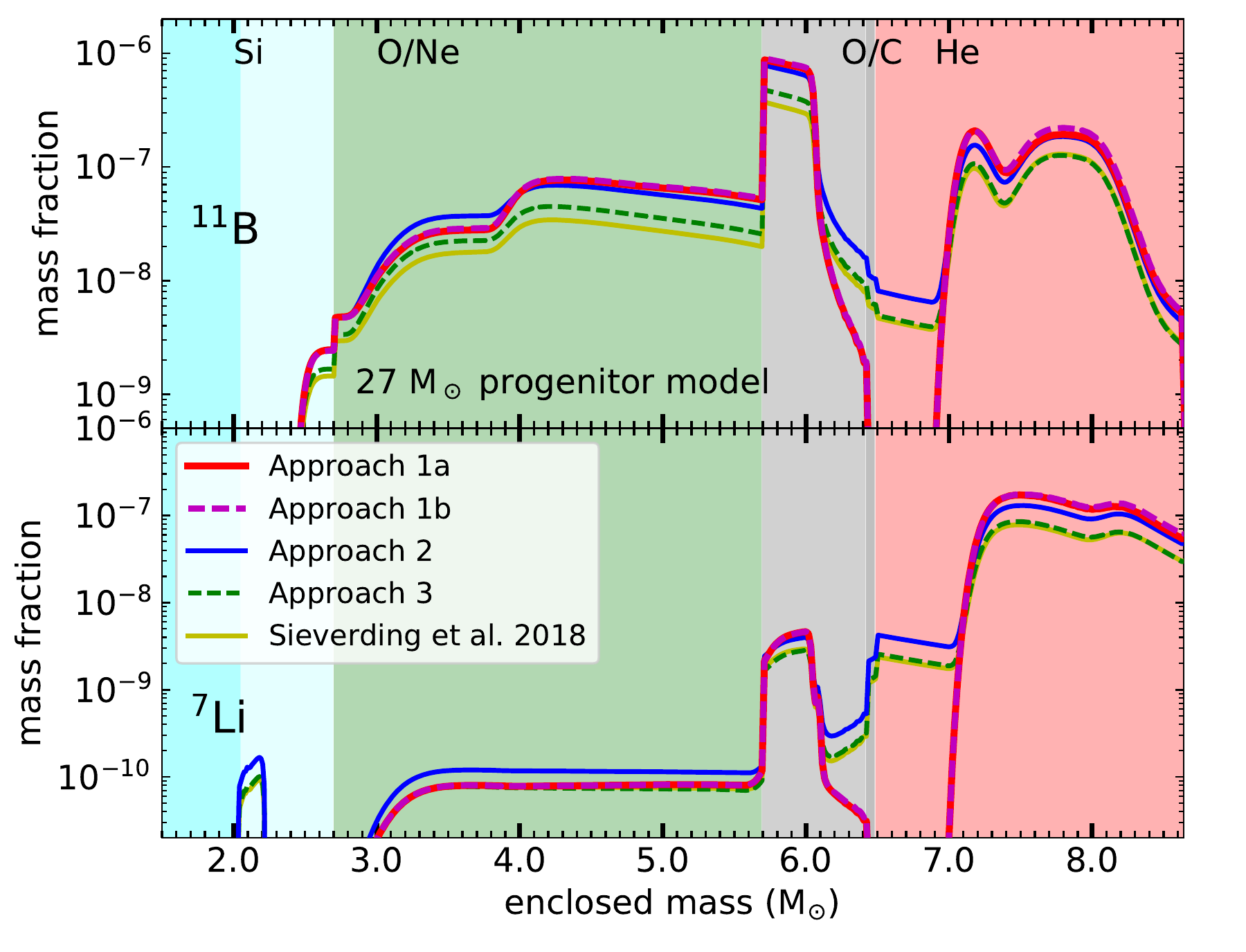}
  \caption{Mass fraction profiles of $^{11}$B (upper panel) and $^7$Li
   (lower panel) for the 27~M$_\odot$ model using different
   descriptions of the neutrino spectra and time evolution. The mass fractions are evaluated after nuclear $\beta$ decays of radioactive isotopes and the
   background colors indicate the compositional shells as indicated at
   the top. Results for our Approaches~1a and~1b are almost identical. Approach~2 shows a reduced mass fraction. For comparison, the
  the results using the low neutrino
   energies from
   \citet{Sieverding.Martinez.ea:2018}, which are similar to our Approach~3, are also shown. \label{fig:b11_signal_compare}}
\end{figure}

In the following, we discuss in detail where in the star the neutrino
nucleosynthesis occurs and which differences result from including the early phases 
of neutrino emission an the time dependence of the average neutrino energies in Approaches~1a and~1b. 

Figure~\ref{fig:b11_signal_compare} shows the mass fraction profiles
of the light nuclides $^{11}$B and $^7$Li as obtained in the
nucleosynthesis studies with our different approaches (see Table \ref{tab:approaches} for an overview of the approaches). Additionally,
we give the results for the same stellar model presented by
\citet{Sieverding.Martinez.ea:2018}, which, as discussed above, are
close to those obtained in our Approach~3.  Both nuclides are produced
in the $\nu$~process by the reaction chains
$^3$H$(\alpha,\gamma)^7$Li$(\alpha,\gamma)^{11}$B and
$^3$He$(\alpha,\gamma)^7$Be$(\alpha,\gamma)^{11}$C following the
neutrino-induced reactions on $^4$He that produce $^3$He and
$^3$H. $^{11}$C is radioactive and decays to $^{11}$B with a halflife
of $20\,\mathrm{min}$. An important contribution to $^{11}$B also comes from the
O/C~shell in which neutrino-induced spallation of $^{12}$C produces
$^{11}$C and $^{11}$B directly.  A detailed discussion can also be
found in \citet{Sieverding.Martinez.ea:2018}.  Due to the inclusion of
the burst and accretion phases, the electron neutrinos have a higher
average energy in our Approach~2 compared to Approach~3 and to the values adopted by
\citet{Sieverding.Martinez.ea:2018}.  As a consequence, the production
of $^{11}$B in the C~shell is increased, because charged-current
reactions contribute almost half to the $^{11}$B
synthesis in this layer. In the He~shell, $^{11}$B is produced from
the spallation products of neutral-current reactions on $^4$He.  As
the average energies of $\nu_x$, $\bar{\nu}_x$ and $\bar{\nu}_{e}$ are
almost the same in our Approach~3 and the study
of~\citet{Sieverding.Martinez.ea:2018} the neutral-current induced
reaction chain produces essentially the same mass fraction of $^{11}$B
in the He~shell. Approach~2, which also uses similar average energies, but a higher luminosity because it includes the early phases of neutrino emission, leads to a slightly larger $^{11}$B abundance in the He~shell. The mass fractions obtained in our Approaches~1a and~1b are very similar to each other, showing that the spectral shape has little impact on the production of these nuclei as discussed in more detail in \S\ref{sec:alphas}.  In both cases, however,
using the time-dependent neutrino energies, the mass fractions turn out to be larger in the
He~and C~shells than found in Approaches~2 and~3. 
The production of $^7$Li in the He~shell stems to a large fraction from
$\nu_{e}$-induced reactions and, for the same reasons as for $^{11}$B,
we observe an increased mass fraction in our Approach~2 compared to Approach~3 and 
the results of \citet{Sieverding.Martinez.ea:2018}. A large fraction of
$^{7}$Li is first produced as $^{7}$Be, started by the
$^4$He$(\nu_{e},e^{-}p)^3$He and $^3$H$(\nu_{e},e^{-})^3$He reactions
followed by an $\alpha$ capture.  Similar to $^{11}$B, our Approaches~1a and~1b
lead to a larger $^7$Li mass fraction than the other calculations.

Using the time-dependent neutrino emission data in Approaches~1a and~1b
increases the total production of $^7$Li and $^{11}$B, but the local mass 
fractions do not increase in all regions of the star.  
The mass fractions of $^{11}$B and
$^{7}$Li in the outer C~shell and at the base of the He~shell at a
mass coordinate of around $6.5\,M_\odot$ are lower for the calculations
with the time-dependent neutrino energies than in the other cases (see
Figure~\ref{fig:b11_signal_compare}). Around
mass coordinate $3.2\,M_\odot$, in the inner O/Ne shell our Approaches~1a and~1b also lead to an
decrease of the mass fractions compared to Approach~2.  These differences
do not affect the total yields noticeably because the mass fractions
in these regions are low compared to those in the He~shell.
Even though the total energy emitted in neutrinos, as defined by
Equation~(\ref{eq:tot_energy}), 
and the number of neutrinos, as defined by
Equation (\ref{eq:tot_number}), already give a good characterization
of the neutrino emission for the $\nu$~process, as can be seen from the 
similar production factors of Approaches~1a/b and~2 in Table~\ref{tab:prodfac_full_signal} , there are subtle
effects that cannot be captured by an averaged approach that reproduces
the same time-integrated energy and number of neutrinos.

To explain these subtle differences, we note that the results of
the $\nu$~process are a competition of neutrino-induced
reactions on one hand and the effect of the shock on the other.  Both
depend on the location in the star where the
competition occurs. In general, at a smaller radius the shock induces
higher temperatures, at which charged particle reactions run
noticeably faster.  Hence, the effectiveness of shock-initiated
nucleosynthesis is increased, but still depends on the available
seed-nuclei. The radius defines also the arrival time of the shock and
which part of the neutrino emission signal acts before arrival of the
shock and which after. The latter point is particularly important to
explain the differences in the neutrino nucleosynthesis yields between
our Approaches~1 and~2, i.e., whether we use the time-dependent
spectra or replace them by spectra with a constant average
energy. Supernova simulations indicate that the average energy of the
emitted neutrinos decreases with time. Taking this into account in our
Approaches~1a and~1b, late neutrinos, emitted after a few seconds, have spectra
shifted to energies lower than those assumed in Approach~2. Thus, in
our Approaches~1a and~1b, neutrino-induced reactions are less effective in
rebuilding the abundances if the shock passage occurred only after a
few seconds.

\begin{figure}[htb]
  \centering
   \includegraphics[width=\linewidth]{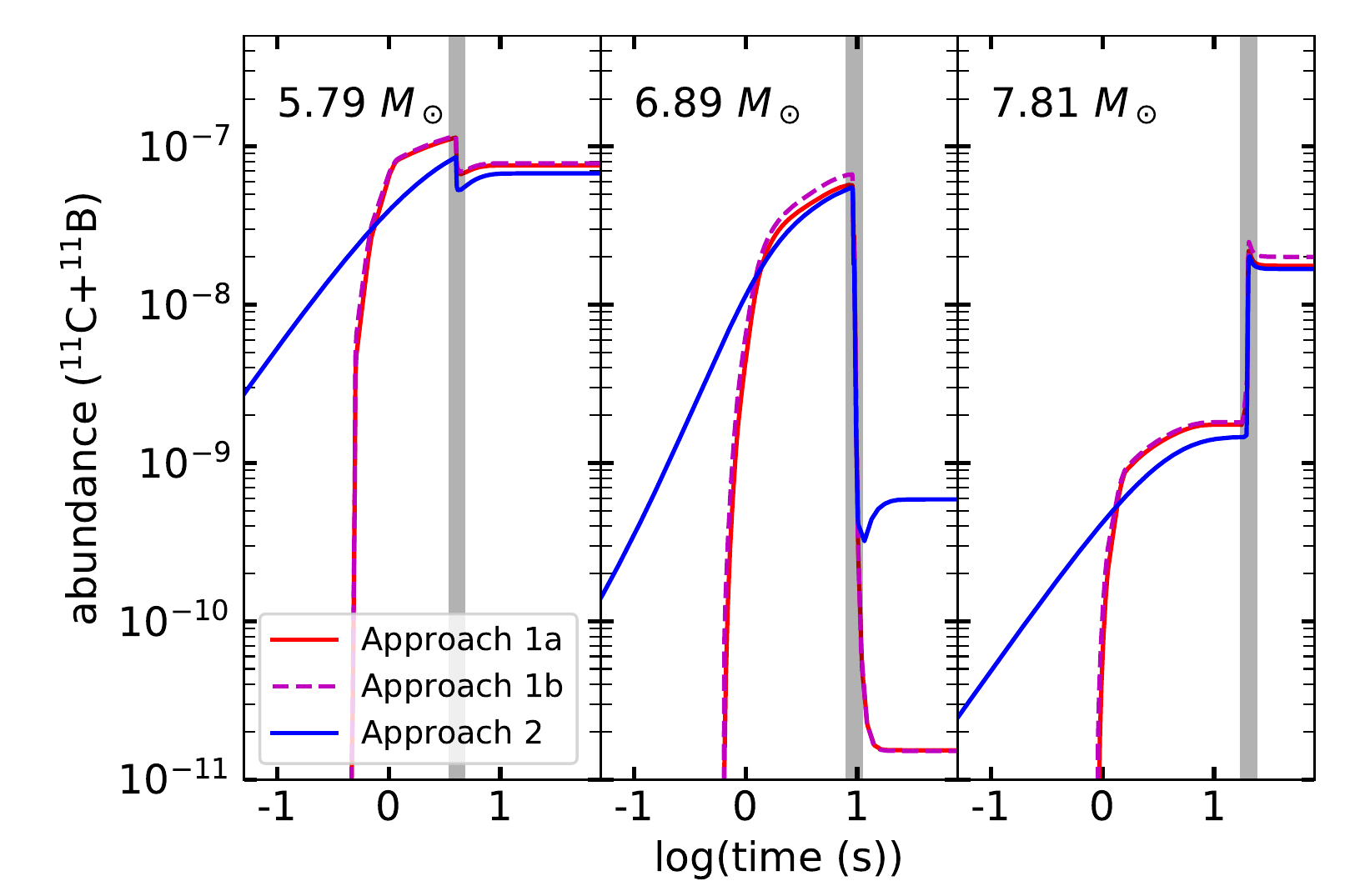}
  \caption{Evolution of the combined abundances of $^{11}$B and
    $^{11}$C for three selected mass zones, corresponding to the
    O/C~shell, the lower He~shell and the middle of the He~shell, where 
    the production is the most efficient. 
    The corresponding mass coordinate is indicated in each panel.  
    The abundance evolution for Approaches~1a and~1b, which both use the time-dependent neutrino energies and luminosities, are shown in comparison to our Approach~2 which uses time-averaged constant values for the neutrino energies. Approach~1b and Approach~2
    both use fixed values of $\alpha=2.3$, whereas Approach~1a uses $\alpha(t)$ based on the supernova simulation. The \textsl{shaded
      vertical line} indicates the time when the supernova shock
    reaches the mass shell. \label{fig:evol_example}}
\end{figure}

Our general conclusions are supported by
Figure~\ref{fig:evol_example}, which compares the abundance evolution
of $^{11}$B as a function of time for our Approaches~1a and~1b to Approach~2. The evolution is almost identical for Approaches~1a and~1b. This shows that the effects of the pinched spectra are much smaller than the effects of including the time-dependence of the neutrino energies, which is neglected in Approach~2 and leads to noticeable differences in the abundance evolution.  
We have depicted three locations in the star, moving
outwards from the left panel to the right panel.  The left panel shows
the evolution in the O/C~shell, where $^{11}$B is made mostly as
$^{11}$C by the $^{12}$C$(\nu,\nu'n)$ reaction.  Using the
time-dependent signal (Approaches~1a and~1b), the production sets in later,
but, due to the higher energy, the production quickly exceeds the case
for the constant average neutrino energy\footnote{The same early
  behavior before shock arrival is also found for the other two mass
  cells, shown in the middle and right panels.}. In the O/C~shell, the
supernova shock reduces the abundance only slightly, because only a
few $\alpha$~particles are available for charged-particle reactions
showing the importance of the composition. The shock passage takes
place around $t=4$~s and matter temperatures reach up to
$1\,\mathrm{GK}$. For both approaches, the neutrino flux is still
substantial after the shock has passed and the abundances recover from
the reduction caused by charged particle reactions
initiated by the shock.  This recovery is slightly stronger in
Approach~2 than in Approach~1b, because the spectra of the late-time neutrinos
have smaller average energies than the constant value adopted
in Approach~2.  The middle panel depicts the situation close to the
base of the He~shell. Here the shock arrives at about $t=10\,\mathrm{s}$ and
reaches a temperature of $0.5\,\mathrm{GK}$. In the He~shell, $\alpha$ particles
are readily available and the shock destroys most of the $^{11}$B
produced before its arrival. At
$t>10$~s the neutrino flux is already too low to recover the $^{11}$B
abundance after the passage of the shock. Moreover, at such late
times, the time-dependent spectra used in Approaches~1a and~1b are noticeably shifted to lower
average energies compared to those adopted in Approach~2. Thus, Approach~2
yields a higher final $^{11}$B abundance than Approaches~1a and~1b at this
location, as in Figure~\ref{fig:b11_signal_compare}. In either case, the final abundances are, however, quite small and do not contribute
noticeably to the total $^{11}$B yield. 
The right panel of
Figure~\ref{fig:evol_example} shows the abundance evolution further
out in the He~shell. The shock arrives after about $t=20$~s and
reaches a maximum temperature of $0.3\mathrm{GK}$. This temperature is too low
to destroy $^{11}$B or $^{11}$C by $\alpha$ reactions, but is still
high enough to initiate the $^{7}$Be$(\alpha,\gamma)$ and
$^{7}$Li$(\alpha,\gamma)$ reactions to produce additional
$^{11}$B. This production also depends on the availability of $^7$Li
and $^7$Be at the time when the shock arrives. Their abundances are
slightly higher in Approaches~1a and~1b than in Approach~2.  In both
approaches, the neutrino flux after passage of the shock is already
too small to further change the $^{11}$B abundance.
At all three locations depicted in Figure \ref{fig:evol_example} we note that the pinched spectra lead to a slight and systematic reduction of the $^{11}$B abundance in calculations with Approach~1a compared to Approach~1b, which assumes $\alpha=2.3$ throughout. These effects are discussed in more detail in \S\ref{sec:alphas}.

\begin{figure}[htb]
   \centering
  \includegraphics[width=\linewidth]{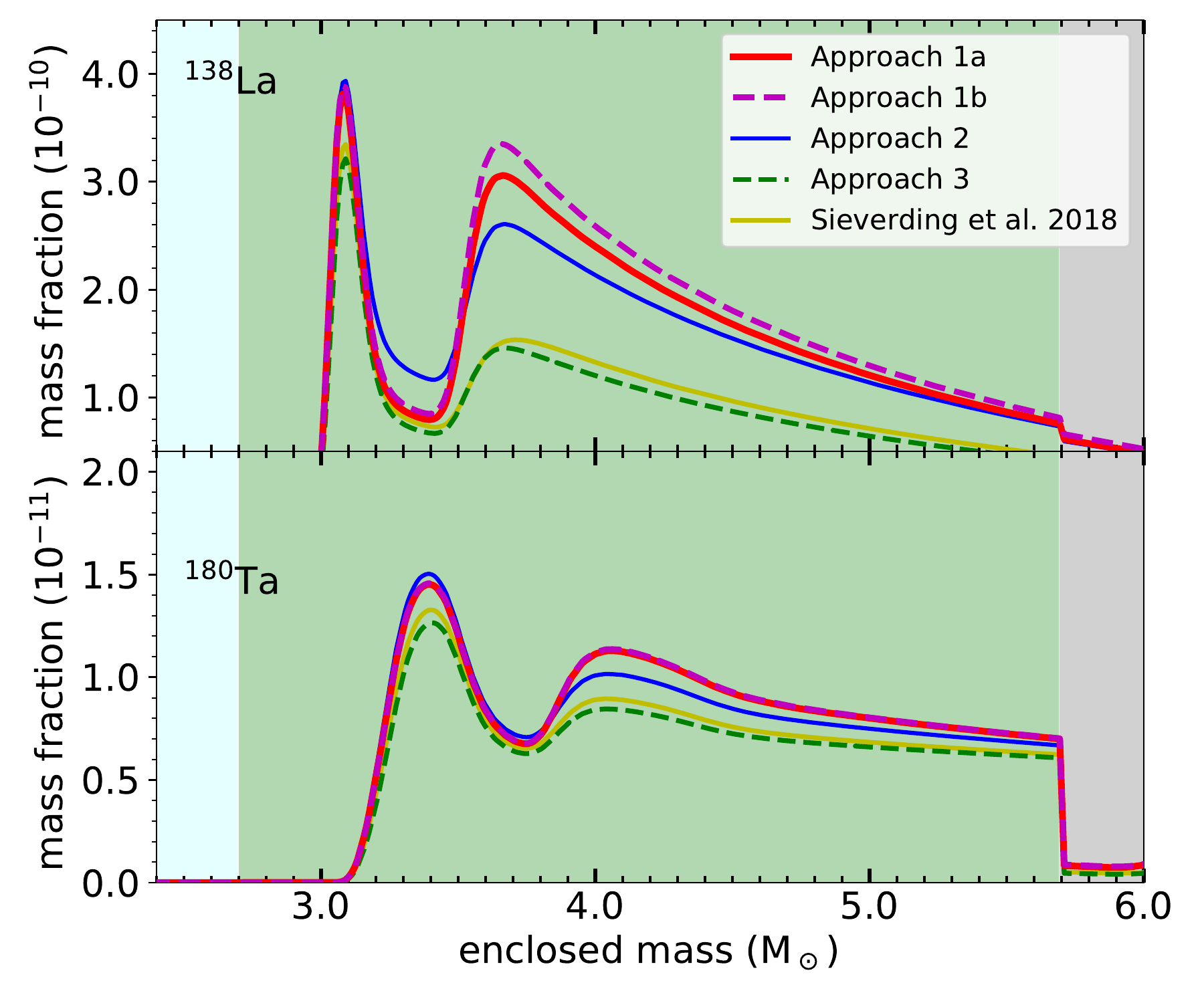}
 \caption{Same as Figure~\ref{fig:b11_signal_compare} but for
   $^{138}$La and $^{180}$Ta from the 27~M$_\odot$ model comparing
   Approaches~1a and~1b to Approach~2. With the full neutrino signal used in
   Approaches~1a and~1b the mass fraction of $^{138}$La is significantly
   increased compared to the approximation of
   Approach~2. Approach~3 results in even lower mass fractions very similar to \citet{Sieverding.Martinez.ea:2018}\label{fig:lata_signal_compare}} 
\end{figure}

Figure~\ref{fig:lata_signal_compare} shows the mass fraction profiles
for $^{138}$La and $^{180}$Ta in the 27~M$_\odot$ star for five 
different treatments of the neutrino emission. Neutrino
nucleosynthesis produces both nuclides mainly in the O/Ne~shell by the
charged-current reactions $^{138}$Ba$(\nu_{e},e^{-})$ and
$^{180}$Hf$(\nu_{e},e^{-})$, respectively. The seed nuclides, $^{138}$Ba
and $^{180}$Hf, which are noticeably more abundant, stem from the
initial composition of the progenitor star modified by the
$\gamma$~process operating prior to the explosion
\citep[see][]{Sieverding.Martinez.ea:2018}. The effect of the shock on the
$^{138}$La and $^{180}$Ta yields results mainly from
a competition of photodissociation, mostly $(\gamma,n)$, and (re-)capture of emitted neutrons.
The matter temperature reached in the O/Ne~shell is
too low to initiate charged particle reactions on the two nuclides due
to their high atomic numbers \citep{Sieverding.Martinez.ea:2018}.  The effect of
photodissociation suggests to distinguish three different regions in
the shell. In the outer shell, the temperatures are low and
photodissociation does not play an important role.  In this region,
$^{138}$La and $^{180}$Ta are then almost unaffected by the supernova
shock and the charged-current reactions induced by the entire $\nu_{e}$
signal add to the final yields.  As discussed above, due to the energy
dependence of the phase space factors early neutrinos with average energies higher than the time-average can already in a short time exceed the production induced by
neutrinos with the constant, time-averaged value for the energies.  As a
consequence, Approaches~1a and~1b give larger mass fractions than Approach~2. Approach~3 and the calculation by \citet{Sieverding.Martinez.ea:2018} result in lower mass fractions due to lower average energies and lower luminosities.
In this region the suppression of the production due to the pinched spectra 
with Approach~1a compared to~1b is most noticeable. 
Closer to the bottom of the O/Ne shell, the mass fractions show a
minimum.  In this region temperatures are sufficiently high during the
shock passage to release neutrons that destroy the  yields of $^{138}$La and
$^{180}$Ta produced before. This is
partly recovered by neutrino nucleosynthesis, where, however, only
late neutrinos contribute.
Similar to the production of $^{11}$B in the lower He~shell discussed above, here Approaches~1a and~1b are less effective in the production of $^{138}$La than
Approach~2, because the average neutrino energies at late times are below
the time-averaged value. The maximum at the bottom of the O/Ne shell
reflects yields which are predominantly produced by the
$\gamma$~process \citep{Sieverding.Martinez.ea:2018}. Here the
contribution of the $\nu$~process is relatively small and arises from
late-time neutrinos, hence the local mass fractions in Approaches~1a and~1b are slightly
smaller than in Approach~2.  
For both
nuclides,$^{138}$La and $^{180}$Ta, the region in the outer O/Ne~shell 
contributes most to the
neutrino nucleosynthesis. Approaches~1a and~1b hence give larger
total yields than Approach~2 (see Table~\ref{tab:prodfac_full_signal}). 
In Approach~3 as well as in \citet{Sieverding.Martinez.ea:2018} the production is also 
noticeably smaller
because the $\nu_{e}$-induced nucleosynthesis is strongly reduced, if the
electron neutrinos emitted during the burst and accretion phases are
not considered.
As discussed in
\citet{Sieverding.Martinez.ea:2018}, the 27~M$_\odot$
progenitor model is special with respect to the $^{180}$Ta production because the yield
predominantly results from a pre-supernova $\gamma$~process. 

\section{Effects of pinched neutrino spectra}
\label{sec:alphas}

\begin{figure}[htb]
  \centering
  \includegraphics[width=\linewidth]{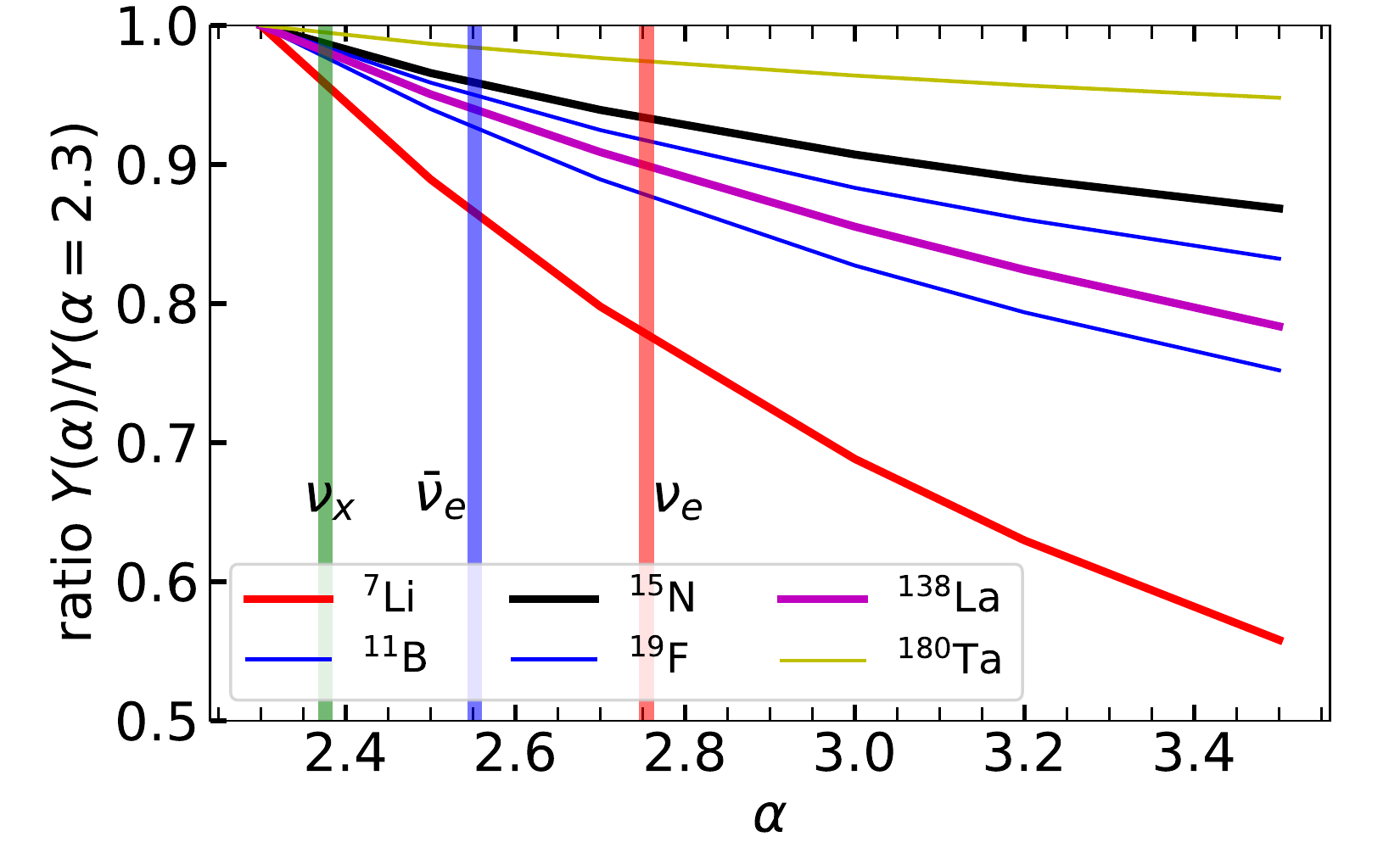}
 \caption{Dependence of production yields on the $\alpha$ parameter of
   the neutrino spectra. For each calculation, the same constant value
   of $\alpha$ was assumed for all neutrino species. The vertical
   lines indicate the average values from the simulations used in the
   previous section.\label{fig:alpha_plot}}
\end{figure}

In the previous section, we have already seen that including 
pinched neutrino spectra in terms of the pinching factor $\alpha(t)$ in Equation~(\ref{eq:alpha_fit}) in our Approach~1a slightly reduces the production of 
the $\nu$-process nuclei compared to Approach~1b, which uses a constant value of~$\alpha=2.3$. 
To illustrate the effect of the spectral pinching on the
nucleosynthesis yields more clearly, we have performed a set of calculations which use 
the time-dependent neutrino luminosities and average energies from the simulation 
but keep the $\alpha$
parameter of the neutrino spectra defined in
Equation~(\ref{eq:alpha_fit}) constant at a range of values. 
Figure~\ref{fig:alpha_plot} shows the
yields that we obtain in the respective nucleosynthesis studies. We
recall that the neutrino-nucleus cross section is reduced for larger
values of $\alpha$, i.e., high energy neutrinos are more suppressed,
and that the reduction is more effective for larger reaction
thresholds. These two observations explain the results observed in
Figure~\ref{fig:alpha_plot}. The light nuclides are affected most as
they involve reactions with $Q$-values of more than 10~MeV. The
production of $^7$Li, involving neutrino-induced reactions on $^4$He
and $^{12}$C with their exceptionally large thresholds, is
particularly sensitive. This can already be seen from the cross
sections in Table~\ref{tab:alpha_reactions}. On the other hand, the
$^{138}$Ba$(\nu_{e},e^{-})$ reaction needs to overcome only a $Q$-value of
around 1~MeV and is therefore less dependent on the high-energy tail
of the neutrino distribution than the neutral-current spallation
reactions.

Whereas the general trend shows that the production of the light
nuclides is most sensitive to the pinching of the neutrino spectra,
the actual calculations using the spectral information from the
supernova simulation give a different result.  This can be seen from
Table~\ref{tab:prodfac_full_signal}, comparing the yields
obtained with the time-dependent, pinched spectra using $\alpha(t)$
from the supernova simulation (Approach~1a) to a calculation which
assumes Fermi-Dirac-line spectra with $\alpha=2.3$ for all neutrino
species, as in previous studies, but also includes the time evolution
of the average energies (Approach~1b).  We find that the effect of
taking into account $\alpha(t)$ is in general rather small and largest for the heavier isotope
$^{138}$La.  The reason can be
understood from the averaged $\alpha$ values, $\langle\alpha\rangle$, that are indicated as vertical lines in Figure~\ref{fig:alpha_plot}
for the different neutrino species. Those values are obtained by taking the time average over
$\alpha(t)$ from the supernova simulation as

\begin{equation}
\langle\alpha\rangle =\frac{1}{t_{\mathrm{end}}-t_0} \int_{t_0}^{t_{\text{end}}} \alpha(t)\,dt ,
\end{equation}
where $t_0$ is the beginning of the neutrino emission around 50~ms
before bounce and $t_{\text{end}}=10$~s, at which time the neutrino
luminosity has dropped below values relevant for nucleosynthesis. We
find that the averaged $\alpha$ for $\nu_x$
($\langle \alpha_{\nu_x}\rangle =2.38$) and $\bar{\nu}_x$
($\langle \alpha_{\bar{\nu}_x}\rangle =2.26$) closely resembles
Fermi-Dirac spectrum, whereas those for electron neutrinos
($\langle \alpha_{\nu_{e}}\rangle =2.75$) and electron antineutrinos
($\langle \alpha_{\bar{\nu}_{e}}\rangle =2.55$) are slightly pinched.
As a consequence of the spectral shape of $\nu_x$ and $\bar{\nu}_x$, the yields of those nuclides that are predominantly
produced by neutral-current reactions change only little if neutrino
spectra from the simulation are used rather than an FD-like
spectrum ($\alpha=2.3$).

Even though the relevant reaction cross section is not very sensitive 
to $\alpha$, the synthesis of $^{138}$La is affected the 
most by the pinching of the neutrino spectra, because it
is produced by $\nu_{e}$-induced charged-current reactions.
Table~\ref{tab:prodfac_full_signal} shows that a pinched neutrino
spectra lead to a 7\% reduction compared to the case of a FD-like
spectra with $\alpha=2.3$. This reduction agrees rather well with the
results shown in Figure~\ref{fig:alpha_plot}, where we find a 9\%
smaller value if we compare the yields calculated for the neutrino
spectra with the averaged value for $\nu_{e}$ neutrinos to the ones for
FD-like spectra\footnote{The small difference results from the fact that the results shown in Figure \ref{fig:alpha_plot} assume constant $\alpha\geq 2.3$ for all neutrino species, whereas Approach~1a uses the values $\alpha(t)$ from the simulation. }. Since $^{180}$Ta is made mainly in the pre-supernova
stage, as discussed above, only the synthesis of $^{138}$La is
affected by the neutrino spectra for this progenitor model, because it is produced by
$\nu_{e}$-induced charged-current reactions.  We stress, however, that
this finding is specific to progenitor models like the one studied
here.  In less massive stars, $^{180}$Ta is also produced via
$\nu_{e}$-induced neutrino reactions and should also be sensitive to the
use of detailed neutrino spectra.

Neutrino oscillations, which we have neglected here, could lead to
much more complicated spectral shapes \citep[see
e.g.,][]{Wu.Qian.ea:2015}, and could also affect the nucleosynthesis
yields.

\section{Sensitivity to the burst luminosity and the accretion time} 
\label{sec:sensitivity}

In this paper, we have for the first time considered the electron
neutrino burst and the standing accretion shock phase of a supernova
explosion in a study of neutrino nucleosynthesis. Although both phases
are firmly established in supernova simulations, some uncertainties
about their properties still remain. In this section, we will explore
what impact these uncertainties have on the neutrino nucleosynthesis
yields.

\begin{figure*}[htb]
  \centering
  \includegraphics[width=0.9\linewidth]{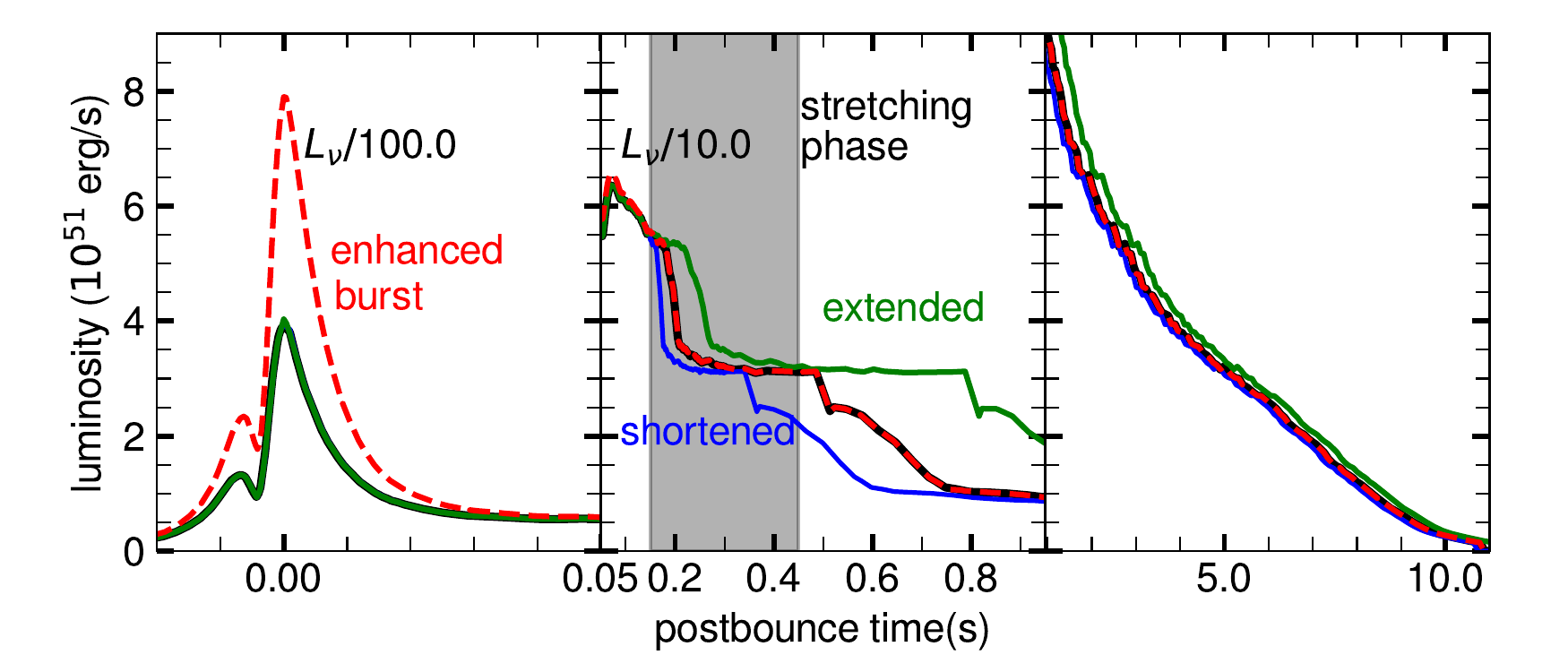}
 \caption{Neutrino luminosities that illustrate the modifications
   applied to the neutrino signal. The shaded region indicates the
   time interval in which the data were stretched in order to achieve
   the modification of the accretion period. \label{fig:illu}}
\end{figure*}

The existence of a $\nu_{e}$ burst is a well-established feature of
supernova simulations that is not very sensitive to the progenitor
model~\citep{Kachelriess.Tomas.ea:2005,Thompson.Burrows.ea:2003}.  The
maximum luminosity of the electron neutrino burst depends, however, on
details of the nuclear equation of state and of the coupling of
neutrinos to hot and dense nuclear matter. This introduces some
uncertainties. For example, studies by~\citet{Marek.Diploma:2003} have shown that the peak luminosity can vary
by around $50\,\%$, typically in the range
$350\text{--}450\times 10^{51}\,\mathrm{erg}\,\mathrm{s}^{-1}$, when different nuclear
equations of state are used.

\begin{figure}[htb]
  \centering
  \includegraphics[width=\linewidth]{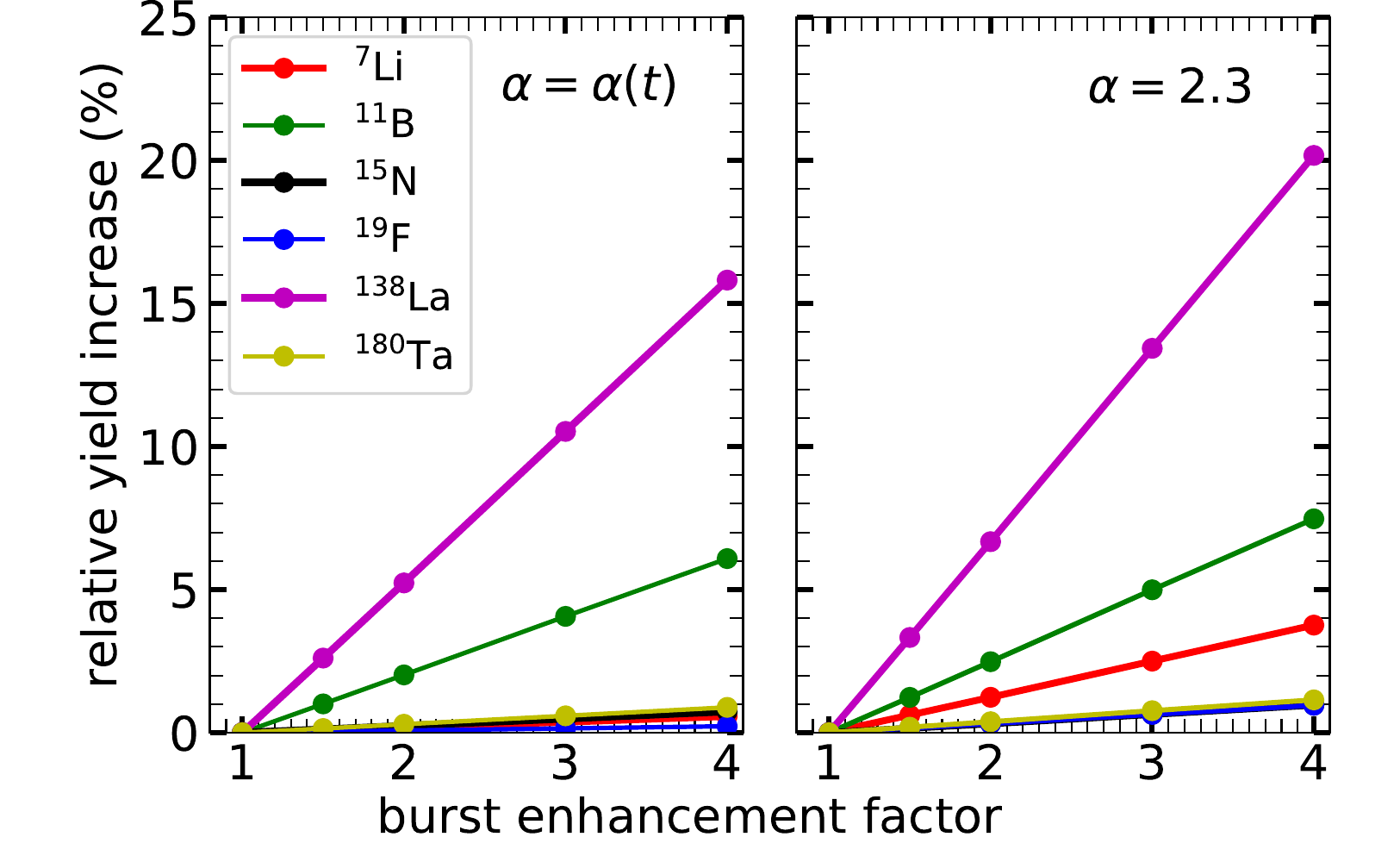}
 \caption{Effect of the variation of the peak luminosity of the
   electron neutrino burst at shock breakout by Equation
   (\ref{eq:burst_enhancement}) for the 27~M$_\odot$ model. The left
   panel shows the results including the effects of pinched neutrino
   spectra and for the results shown in the right panel a constant
   value of $\alpha=2.3$ has been assumed, approximating pure FD
   spectra. The pinching reduces the effect of the $\nu_{e}$ burst
   enhancement.\label{fig:burst_effects}}
\end{figure}

We study the impact of this uncertainty by varying the neutrino burst peak
luminosity. To this end, we fold the numerical electron neutrino
luminosities $L_{\nu_{e}}^0(t)$ with a Lorentzian centered around the
time of maximum luminosity, $t_{\text{burst}}$; i.e., we assume
\begin{equation}
\label{eq:burst_enhancement}
  L_{\nu_{e}}(t)=L_{\nu_{e}}^0(t) \, \left[ 1+ (A-1)\frac{w^2}{(t-t_{\text{burst}})^2+w^2) } \right],
\end{equation}
with a half-width, $w$, of $12.5\,\mathrm{ms}$ and a variable enhancement factor,
$A$. Using this ansatz, the enhancement factor directly translates
into an increase of the peak luminosity by a factor of $A>1$.  The width of
the enhancement is chosen such that the effect is limited to the
duration of the neutrino burst.
The modification is illustrated by the red dashed line in
Figure~\ref{fig:illu}. In our calculation, we assume that neither the
luminosities of the other neutrino species nor the neutrino spectra
are modified by the variation of the burst luminosity.
Figure~\ref{fig:burst_effects} shows the relative yields of the
$\nu$-process nuclides as a function of the enhancement factor, $A$.
Even though we do not expect to find variations of the peak
luminosities in supernova simulations by more than 50\%, we vary $A$
by up to a factor $4$ for illustrative purposes. The yields increase
by enhancing the burst luminosity, and in the regime we study, this is
a linear effect, that overall, however, is small.
The neutrino burst has an impact on the yields of nuclides that are
produced by $\nu_{e}$-induced charged-current reactions. Thus, we
observe the largest changes due to variations of the burst luminosity
for $^{138}$La.  For progenitors with different masses, we expect the $^{180}$Ta
yield to show a similar dependence. We note again that the current
27~M$_\odot$ progenitor model represents an exception for the
$^{180}$Ta production, as it is synthesized by
a pre-explosive 
$\gamma$~process \citep[e.g.,][]{Rauscher.Heger.ea:2002} and not by $\nu_{e}$-induced charged-current reactions,
as for other progenitor stars \citep{Sieverding.Martinez.ea:2018}.
Consequently, the $^{180}$Ta yield in Figure~\ref{fig:burst_effects}
shows very little variation when the burst luminosity is increased.
The effects of the variation of the burst luminosity on
the yields of $^7$Li, $^{15}$N, and $^{19}$F are negligible, because these isotopes 
are mainly produced
by neutral-current reactions. 
Half of the $^{11}$B production in the C/O~shell, which constitutes a major part of the total yield, results from charged-current
reactions; hence, $^{11}$B shows some sensitivity to the variation of the
burst luminosity.
The electron neutrino spectra are strongly pinched during the neutrino
burst (see Figure~\ref{fig:sim_signal_SFHo}), i.e., the high energy tail
is significantly reduced compared to an FD-like spectrum with
$\alpha=2.3$. As a consequence the impact of the neutrino burst on the
neutrino nucleosynthesis yields is noticeably reduced, if---like in
the present work---  pinched neutrino spectra are
considered rather than Fermi-Dirac spectra. This can be seen by
comparing the left panel of Figure~\ref{fig:burst_effects}, which is
calculated with the time-dependent $\alpha(t)$, to the right panel, where
FD-like spectra have been assumed.

Another qualitatively well-established, but quantitatively uncertain
feature of the supernova neutrino emission is the accretion phase.
\citet{Mueller:2015} and \citet{Bruenn.Lentz.ea:2016} have reported
simulations in which accretion-driven neutrino emission can persist
for up to $\sim 1$~s after core bounce.  After a successful shock
revival, the mass accretion rate drops substantially also in multi-D
simulations and the neutrino emission is reduced.  Most modern
simulations predict a delay time until shock revival of a few hundred
milliseconds after core bounce.  The duration of this phase of accretion
until the onset of the explosion is subject to the treatment of
multidimensional turbulence that might not be fully numerically
resolved yet~\citep{Radice.Ott.ea:2016}, even in state-of-the-art
three-dimensional simulations. The duration is also sensitive to the
neutrino interactions in nuclear matter~\citep{OConnor.Couch:2018,Burrows.Vartanyan.ea:2018}.

\begin{figure}[htb]
  \centering
  \includegraphics[width=\linewidth]{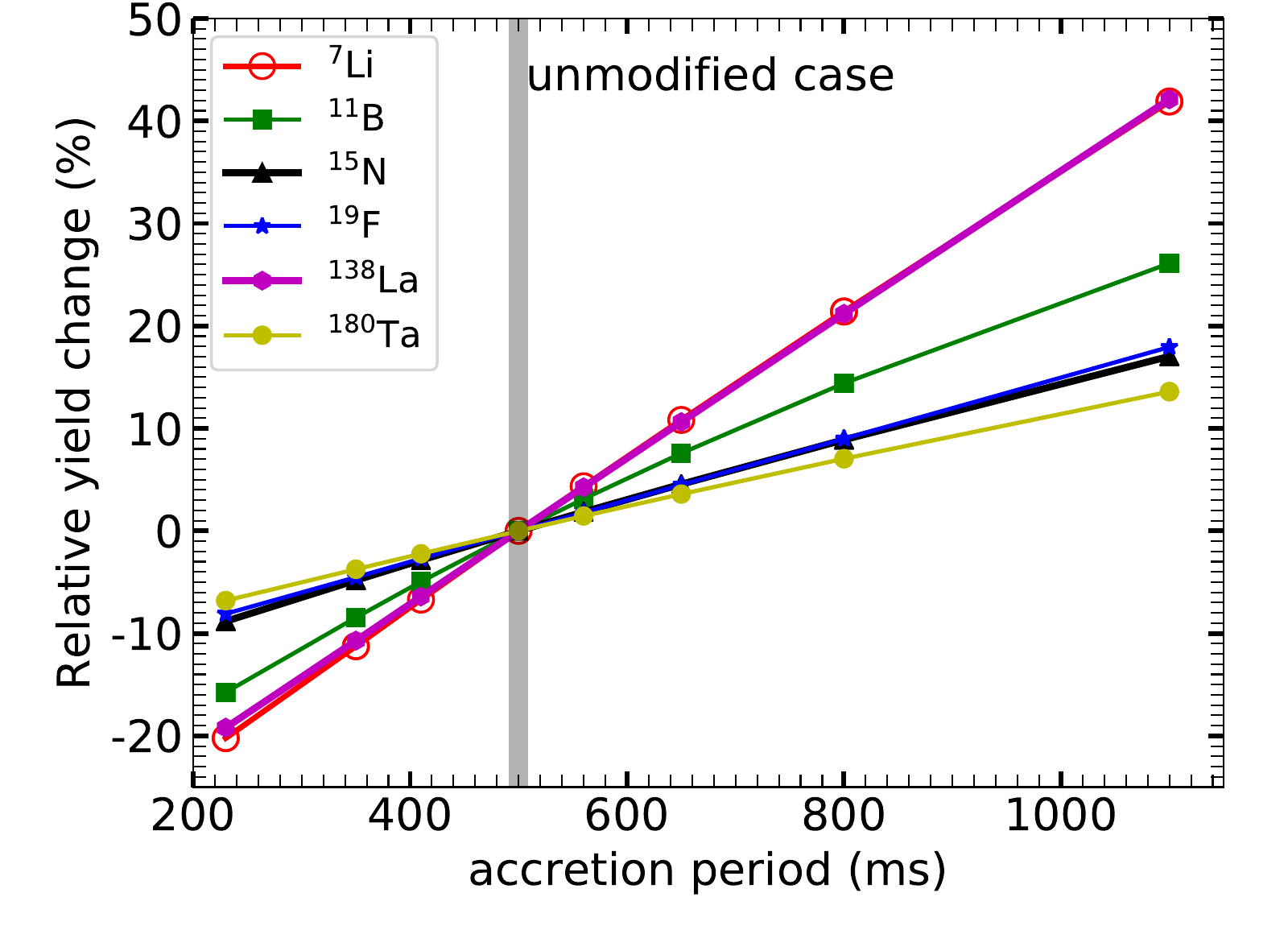}
  \caption{Relative change of the production of $\nu$-process nuclei
    due to the variation of the duration of the accretion phase as
    illustrated in Figure~\ref{fig:illu}. The gray vertical line
    indicates the result for the original, unmodified neutrino
    data. The production factor of $^{138}$La exceeds a production
    factor of unity for an accretion period of
    $1100\,\mathrm{ms}$. \label{fig:acc_effects}}
\end{figure}

Here we explore the sensitivity of the $\nu$-process nucleosynthesis
to the delay time until shock revival and thus the duration of the
phase of most vigorous accretion. In order to do so, we have modified
the neutrino signal shown in Figure~\ref{fig:sim_signal_SFHo}, which
has been the default of our calculations so far (Approach~1a), by
stretching the time steps in the interval between $0.15\,\mathrm{s}$ and 0.45~s to
vary the length of the accretion phase. The same time stretching has
been applied to the neutrino energies and $\alpha(t)$. In practice, we
have replaced each value $t_i$ of the temporal grid of the simulation
dumps by $t_i'= t_{i-1}'+f_{\text{s}}\, (t_i-t_{i-1})$, where $i$ runs
over the simulation dumps with $0.15\ \text{s}<t_i<0.45\ \text{s}$ and
$f_{\text{s}}$ is a stretching factor. Outside the stretching interval, 
the timesteps are left unchanged with an offset to compensate for the modifications. 
This is also illustrated in
Figure~\ref{fig:illu} by the blue and green lines.  In this way, we
can mimic a variation of the delay time without making any further
assumptions about the structure of the signal. Physically, this
variation can be associated with a modification of the mass accretion
rate, that is sensitive to the complex dynamics of multidimensional
fluid flows.  Figure~\ref{fig:acc_effects} shows the enhancement of
the yields of the $\nu$-process nuclides as function of the modified
duration of the accretion phase.  The results obtained for an
accretion period of $500\,\mathrm{ms}$ as adopted throughout this paper so far, is
indicated by the gray vertical bar. The yields grow with increasing
length of the accretion phase. During the accretion phase the $\nu_{e}$
luminosity is noticeably larger than the $\nu_x$ luminosity and, on
average, the $\nu_{e}$ spectra are shifted to higher energies compared
to the later phases. This shift is larger for $\nu_{e}$ than for the other neutrino species. Both effects favor charged-current over
neutral-current reactions. As a consequence, the production of
$^{138}$La can be noticeably enhanced if the accretion phase lasts
longer. (For the reasons discussed above, we expect a similar effect
for $^{180}$Ta in other progenitor stars.) This effect is even
underestimated in the current treatment, because
$\langle E_{\nu_{e}} \rangle$ and $\langle E_{\bar{\nu}_{e}} \rangle $ are
expected to continue to rise during a longer accretion period in
self-consistent models. The yield of $^7$Li also grows noticeably in relative terms
with the length of the accretion phase. This is related to the 
fact that $^7$Li is mostly produced at large radii in the He~shell, where the shock heating 
is not significant and the whole neutrino signal can come to bear.
For both, $^7$Li and $^{11}$B charge-current reactions, which are the most enhanced by an extended accretion period, contribute about $50\,\%$ of the total production. The production $^{11}$B in the O/C~shell, however, is less sensitive to 
the early neutrino emission, because of the stronger shock heating (see Figure \ref{fig:evol_example}).
In general, the yields for nuclides which are
mainly produced by neutral-current reactions, are less
sensitive to the duration of the accretion phase.
The electron flavor neutrinos also induce particle emission by
neutral-current scattering. However, due to their lower energies their
contribution is typically small compared to the heavy flavor
neutrinos. For example, the charged-current reactions
$(\nu_{e},e^{-} p)$ and $(\bar{\nu}_{e},e^{+} n)$ on $^{20}$Ne and
$^{16}$O also contribute to the yields of $^{19}$F and $^{15}$N, respectively, but their production is
increased by less than 10\% even if the accretion period lasts
for more than 1~s.
A shortened accretion phase reduces the yields of $^{7}$Li and
$^{138}$La  
substantially. A reduction of the delay time of the explosion to around $200\,\mathrm{ms}$ 
would reduce the production factor of $^{138}$La normalized to $^{16}$O down to $0.5$.
Assuming that supernovae are the main production site of $^{138}$La, very fast or even prompt supernova explosions would make it more difficult to explain the solar 
abundance of $^{138}$La.
A prolonged accretion phase, on the other hand, selectively increases the production of 
the heavier elements without risking an overproduction of the light
elements $^{7}$Li, $^{11}$B, $^{15}$N, and $^{19}$F.  
Recent studies by \citet{Travaglio.Roepke.ea:2011} and
\citet{Travaglio.Rauscher.ea:2018} suggest that a substantial fraction
of $^{180}$Ta could also come from the $\gamma$~process in Type Ia
supernovae with strong, accretion-induced $s$-process enrichment, but
they do not find a significant contribution to $^{138}$La.  This
leaves $^{138}$La as a good candidate that could be a clear indicator
of the $\nu$~process in core-collapse supernovae.  We note that the
duration of the accretion phase can noticeably enhance the production
factor of $^{138}$La (and, as we expect, for $^{180}$Ta for other
progenitor stars).  As both yields are close to the solar production
factors, an additional production due to an extended accretion phase
might generate some tension due to overproduction.
For the $27\,M_\odot$ model studied here, the production factor of
$^{138}$La exceeds unity only for an extremely long accretion period
of 1100~ms. Such a long delay time is currently not predicted by
simulations, making them consistent with the solar $^{138}$La abundance.

More quantitative statements, however, require extending our study to
the full range of supernova progenitor models with a consistent
treatment of the neutrino emission.

\section{Conclusions}

We have performed neutrino nucleosynthesis calculations for a $27\,M_\odot$ stellar progenitor model that, for the first time, use time-dependent neutrino
emission spectra as obtained from modern (one-dimensional) supernova
simulations. In particular, our approach includes the neutrino
emission during the early electron neutrino burst and accretion phases
of the explosion. Furthermore, we use the spectral form of emitted
neutrinos as predicted by the simulations, and hence account for their
deviations from a zero degeneracy Fermi-Dirac spectrum, which has been
the default assumption in previous studies of $\nu$-process
nucleosynthesis.

Our calculations confirm that selected nuclides ($^7$Li, $^{11}$B,
$^{15}$N, $^{19}$F, $^{138}$La, and $^{180}$Ta) are partly or
predominantly produced by the $\nu$~process.  The production of the
nuclei $^{138}$La and $^{180}$Ta is mainly due to $\nu_{e}$-induced
charged-current reactions, whereas neutral-current reactions, induced
mainly by the neutrino species other than electron neutrinos due to
their higher average energies, contribute to the $\nu$-process yields
of the other four nuclides. We find that our calculation with
time-dependent neutrino emission spectra results in noticeably higher
yields than obtained in the spirit of previous approaches,
i.e., assuming constant neutrino average energies appropriate for the
neutrino emission from the proto-neutron star cooling phase.  In an
additional approach, we have shown that the yields obtained with the
fully time-dependent neutrino emission can be reproduced within a few
percent if the constant average energy takes into account the full
neutrino emission including the neutrino burst and accretion phases.

We have found that the electron neutrino burst gives a rather small
contribution to the total $\nu$-process yields, even for those
nuclides produced by $\nu_{e}$ charged-current reactions. Neutrino
emission during the accretion phase has a larger impact on the total
yields, enhancing the production between about $5\,\%$ for nuclides
made by neutral-current reactions and nearly $20\,\%$ for those with
strong charged-current contributions, caused by the relatively high
luminosities and average energies of electron neutrinos during the
accretion phase. In this paper we used the neutrino emission
spectra from the supernova simulation of
\citet{Mirizzi.Tamborra.ea:2016}, in which the accretion phase lasts
for around 500~ms. There is, however, some uncertainty
about the duration of the accretion period. We have shown that the
production of $^{138}$La increases by more than $40\,\%$ 
if vigorous
accretion-induced neutrino emission lasts for one second.

The calculated neutrino spectra deviate from a Fermi-Dirac
distribution. This deviation is quite strong during the burst phase
where the electron neutrino spectra are strongly pinched, i.e., shifted
to smaller energies. This is one reason why the electron neutrinos
emitted during the burst have a small impact on the $\nu$-process
yields. Overall, the consideration of pinched spectra has a negligible
effect on the neutrino nucleosynthesis results. The main reason is
that the spectra during the cooling phase are well approximated by a
Fermi-Dirac distribution.

We have shown that the outcome of neutrino nucleosynthesis in general,
and of our improved time-dependent treatment of neutrino emission in
particular, is a subtle competition of neutrino-induced reactions and
the effect of the shock wave. The competition depends sensitively on
the radial position in the star at which the nucleosynthesis occurs.  As the
neutrinos travel faster than the shock, parts of the neutrino
nucleosynthesis has already happened when the shock arrives. The shock
can destroy this abundance if the associated temperatures are high
enough. Importantly, the radial position in the star, i.e., the time at which
the shock operates, also decides which temporal portion of the
neutrino emission can induce nucleosynthesis reactions after the
passage of the shock. As later emitted neutrinos have smaller average
energies, their contributions to the $\nu$-process yields is reduced,
in particular compared to studies which assume constant neutrino
average energies.

In general, we have demonstrated that a proper treatment of the
time-dependent neutrino emission has a significant impact on the
neutrino nucleosynthesis yields. Such a treatment is hence
indispensable if one wants to use neutrino nucleosynthesis as a
thermometer for supernova neutrino emission\citep[e.g.,][]{Heger.Kolbe.ea:2005}, or even, to constrain
neutrino properties like mass hierarchy of mixing angles
\citep[see][]{Yoshida.Kajino.ea:2005,Yoshida.Kajino.b.ea:2006,Kajino.Mathews.ea:2014}.
Yet, we also see some necessary next steps. Our calculation has
been restricted to a $27\,M_\odot$ progenitor star, consistent with the 
supernova simulation (which, however, has quite
exceptional features for the $\nu$-process production of
$^{180}$Ta). Additional calculations for the broad range of
progenitors which contribute to the galactic chemical evolution are
desirable.  At this point we have also neglected neutrino flavor
conversion effects on the $\nu$~process.  Finally, our study is based
on the neutrino emission data obtained from a one-dimensional
supernova simulation. Thus, it does not properly describe effects on
the neutrino emission which are caused or influenced by
multidimensional phenomena. Such improvements, which are particularly
relevant for the accretion phase, will be the subject of future
studies.

\begin{acknowledgments}
  This work was supported in part by the US Department of Energy
  [DE-FG02-87ER40328 (UM)].  GMP is partly supported by the Deutsche
  Forschungsgemeinschaft (DFG, German Research Foundation) -
  Projektnummer 279384907 - SFB~1245 ``Nuclei: From Fundamental
  Interactions to Structure and Stars'' and the ``ChETEC'' COST Action
  (CA16117), funded by COST (European Cooperation in Science and
  Technology).  At Garching, this project was supported by the
  European Research Council through grant ERC-AdG No.\
  341157-COCO2CASA, and by the Deutsche Forschungsgemeinschaft through
  Sonderforschungbereich SFB 1258 ``Neutrinos and Dark Matter in Astro-
  and Particle Physics'' (NDM) and the Excellence Cluster Universe (EXC
  153; \url{http://www.universe-cluster.de/)}.  The neutrino data of
  RB's simulation is available through the core-collapse supernova
  archive
  \url{https://wwwmpa.mpa-garching.mpg.de/ccsnarchive/archive.html}.
  AH was supported by an Australian Research Council (ARC) Future
  Fellowship (FT120100363), by the US National Science Foundation
  under Grant No. PHY-1430152 (JINA Center for the Evolution of the
  Elements and by TDLI/SJTU though a grant from Science and Technology
  Commission of Shanghai Municipality (grant No. 16DZ2260200) and
  National Natural Science Foundation of China (grant No.11655002).
 \end{acknowledgments}
 \newpage
 \appendix
 \section{Selected Neutrino-Nucleus Cross Sections as a Function of Incident Neutrino Energy}
 \label{sec:appendix}
In our calculations for Approach~1a, we have performed the folding of the neutrino-nucleus cross sections $\sigma_\nu(E_\nu)$ with the
neutrino spetrum $n_\nu(E_\nu,\langle E_\nu \rangle(t), \alpha(t))$ according to Equation~(\ref{eq:integ_csect}) at every time $t$
for the selected set of reactions shown in Table \ref{tab:alpha_reactions}. The values of $\sigma_\nu(E_\nu)$ for reactions on $^4$He and
$^{12}$C have been taken from shell model calculations by \citet{Yoshida.Suzuki.ea:2008}\footnote{Machine readable tables of the cross sections
are available as online-only material \url{https://iopscience.iop.org/article/10.1086/591266/fulltext/}.}. The other reactions follow
\citet{Sieverding.Martinez.ea:2018} and have been published only as spectrally averaged values.
The cross sections are shown in Figure 10 as a function of the incident neutrino energy, and are provided as numerical values in supplemental material to this article
For the production of $^{15}$N, neutral-current reactions on  $^{16}$O that lead to the emission of a proton or a neutron are the most relevant and require an incoming neutrino with an energy of at least $12\,\mathrm{MeV}$. The two-particle emission processes $^{16}$O$(\nu,\nu' \alpha p)^{11}$B and $^{16}$O$(\nu,\nu' \alpha n)^{11}$C contribute slightly to the production of $^{11}$B and are only possible for neutrinos with sufficiently high energies. Shell-model calculations for neutrino reactions on $^{16}$O have recently been presented by \citet{Suzuki.Chiba.ea:2018} and found significantly larger cross sections for $^{16}$O$(\nu,\nu' \alpha p/n)$.  
The production of $^{19}$F by neutral-current reactions on  $^{20}$Ne that lead to particle emission also requires to overcome a separation energy of $12.84\,\mathrm{MeV}$ for protons and $16.84\,\mathrm{MeV}$ for neutrons.
 The reaction cross sections on $^{20}$Ne that we use here take into
account experimental data from \citet{Anderson.Tamimi.ea:1991} as in \citet{Heger.Kolbe.ea:2005}. The cross-sections for reactions on $^{138}$Ba and
on $^{180}$Hf are based on measurements of the Gamow-Teller strength by \citet{Byelikov.Adachi.ea:2007}. The transitions to 
$^{138}$La and on $^{180}$Ta without particle emission have a low threshold of  $1.05\,\mathrm{MeV}$ and $0.70\,\mathrm{MeV}$
respectively. Hence, the cross-section reaches already high values at relativey low neutrino energies. The neutron separation energies are  
$7.45\,\mathrm{MeV}$ for $^{138}$La and  $6.65\,\mathrm{MeV}$ for $^{180}$Ta. For neutrinos with energies larger than $30-40\,
\mathrm{MeV}$ the neutron emission dominates over the transition to the isobar. Neutron captures on $^{137}$La produced by $^{138}$Ba$(\nu_e,e^- n)^{137}$La contribute up to $10\,\%$ of the total $^{138}$La production \citep{Sieverding.Martinez.ea:2018}. At an incident neutrino energy of $65\,\mathrm{MeV}$ the 2-neutron emission channel beomes dominant. Due to the high charge of these nuclei, the emission of protons is strongly suppressed by the Coulomb barrier for reactions on $^{138}$Ba and $^{180}$Hf.

 \begin{figure}
 \centering
  \includegraphics[width=\linewidth]{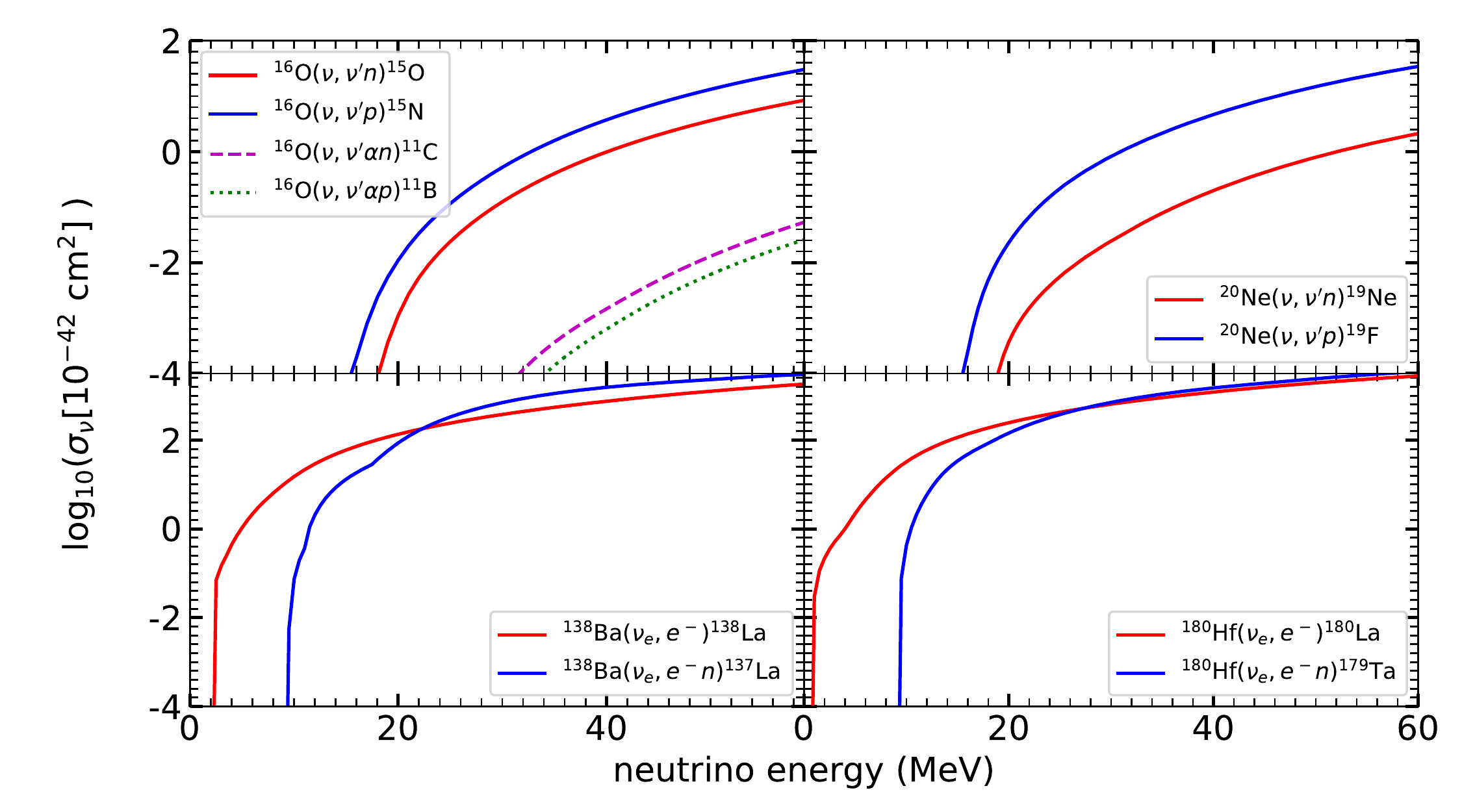}
  \caption{\label{fig:csect}Energy-dependent cross sections for neutrino nucleus reactions for which the effect of pinched neutrino spectra has been included in Approach~1a.}
 \end{figure}
\newpage

\bibliographystyle{apjnew}

\end{document}